\documentclass[aps,prl,twocolumn,
amsmath,amssymb,nofootinbib,superscriptaddress,floatfix]{revtex4}
\usepackage{amsmath}
\usepackage{amssymb}
\usepackage[pdftex]{color}
\usepackage{graphicx}
\usepackage{dcolumn} 
\usepackage{bm} 
\usepackage{longtable}
\usepackage{ulem}

\usepackage{color}
\usepackage{graphicx}
\usepackage{curves}
\usepackage{epic}
\usepackage{wasysym}
\usepackage{epsf, times}
\usepackage{subfigure}
\usepackage{amsthm}

\DeclareRobustCommand{\SkipTocEntry}[4]{}

\newcommand{\bs}[1]{{\boldsymbol{#1}}}

\begin{document}

\title{Majorana zero modes on a necklace 
      }

\author{Jian Li} 
\affiliation{
Department of Physics,
Princeton University,
Princeton, NJ 08544, USA
            } 

\author{Titus Neupert} 
\affiliation{
Princeton Center for Theoretical Science,
Princeton University,
Princeton, NJ 08544, USA
            }            

\author{B. Andrei Bernevig} 
\affiliation{
Department of Physics,
Princeton University,
Princeton, NJ 08544, USA
            } 
            
\author{Ali Yazdani} 
\affiliation{
Department of Physics,
Princeton University,
Princeton, NJ 08544, USA
            }

\begin{abstract}

\end{abstract}

\date{\today}

\maketitle

\textbf{
Non-Abelian quasiparticles have been predicted to exist in a variety of condensed matter systems. 
Their defining property is that an adiabatic braid between two of them results in a  nontrivial change of the quantum state of the system. To date, no experimental platform has reached the desired control over non-Abelian quasiparticles to demonstrate this remarkable property.
The simplest non-Abelian quasiparticles -- the Majorana bound states (MBS) -- can occur in one-dimensional (1D) electronic nano-structures proximity-coupled to a bulk superconductor -- a platform that is being currently explored in great depth both theoretically~\cite{Kitaev01,Fu08,Lutchyn10, Oreg10, Sau10,Alicea10,Potter10, Alicea11,Halperin12,Stanescu13} and experimentally~\cite{Mourik12,Das12,Deng12,Rokhinson12,Churchill13}.
When tuned appropriately, such nano-wires can localize MBS at their ends, a pair of which forms a two-level system that is robust to local perturbations. This constitutes a topologically protected qubit that can serve as the building block for a topological quantum computer~\cite{Moore91,Read00,Ivanov01,Nayak08,Alicea12}. 
To implement braiding operations among the MBS, schemes that allow to move MBS across wire networks have been explored theoretically~\cite{Alicea11, Hassler11}. 
Here, we propose a simpler alternative setup, based on chains of magnetic adatoms on the surface of a thin-film superconductor, in which the control over an externally applied magnetic field suffices to create and manipulate MBS. We consider specific one-dimensional patterns of adatoms, which can be engineered with scanning-tunneling-microscope based lateral atomic manipulation techniques~\cite{Crommie93,Nilius02,Floesch04}, and show that 
they allow for the creation, annihilation, adiabatic motion, and braiding of pairs of MBS by 
varying the magnitude and orientation of the external magnetic field.}

One venue to a 1D topological superconductor that supports MBS are chains of localized magnetic moments, which are coupled to superconducting electrons~\cite{Choy11, Nadj-Perge13, Nakosai13, Klinovaja13, Braunecker13, Vazifeh13, Pientka13a, Po13, Pientka13b}. Along the chain, Shiba bound states~\cite{Yu65,Shiba68,Rusinov68,Balatsky06,Yazdani97,Ji08} form in the superconducting gap. The superconducting electrons mediate Ruderman-Kittel-Kasuya-Yoshida (RKKY) interactions between the localized magnetic moments of the chain. This mechanism evokes a spiral magnetic order, where the pitch is given by either twice the Fermi momentum of the band of Shiba states if the RKKY interaction is 1D~\cite{Kim14} or the Fermi surface of the underlying superconductor is largely one-dimensional, or by the spin-orbit coupling if the RKKY interaction is 2D \cite{Kim14}. In the anisotropic case,  the back-action of the magnetic order on the conduction electrons lifts their degeneracy by opening a gap exactly at the Fermi energy for one spin orientation~\cite{Choy11,Braunecker09,Braunecker10,Meng13,Scheller13}. This effect requires no fine-tuning of parameters and occurs also in superconducting systems~\cite{Klinovaja13}, rendering them a topological 1D superconductor. In the more realistic 2D situation \cite{Kim14}, the existence and pitch of the helix depend on the amount of positional disorder and on the spin-orbit coupling interaction of the surface.  
 
Here, we show theoretically that such helical magnetic chains, assumed to have formed a helix as the experimental situation suggests \cite{Menzel13}, on the surface of a thin film superconductor, can be used to create and manipulate MBS in a simple way by changing, respectively, the amplitude and orientation of an external magnetic field. Scanning-tunneling-microscope (STM) tips can be used to control each magnetic atom's position individually to create any desired geometrical network of magnetic adatoms on the surface of a superconductor. STM can be further used to characterize the magnetic texture of these networks (using spin-polarized tips) as well as presence of MBS in a spatially resolved manner. In an applied external magnetic field of appropriately chosen strength, we find that a circular ad-atom chain features two trivial and two topological superconducting segments, whose interfaces each host a single MBS. The position of the MBS depends on the orientation of the external magnetic field in the plane of the superconducting surface. Rotating the magnetic field at an appropriate rate thus can adiabatically move the MBS along the circle. Based on this mechanism, we further propose structures that implement braiding and a $\sigma_x$ gate operation on qubits formed from the MBS, via rotating the external magnetic field by $2\pi$. All operations can be performed with the magnetic field in the plane of the superconducting surface, leaving thin-film superconductivity intact. We stress that, while the helix forming mechanism is currently under theoretical debate, the setups that we propose are not limited to systems with helical order. They can also be realized in nano-wires with Rashba spin-orbit coupling (SOC) and ferromagnetic order (see Supplemental Material~\cite{Supp}, Sec.~B~2).

\textit{Magnetic structure of the chain} --- 
The magnetic order of the chain and its origin is a subject of intense scrutiny. The situation is likely to become even more theoretically complicated in the presence of an externally applied magnetic field which can possibly influence the formation and structure of the spin helix. In the absence of ab-initio calculations, we give a scenario of one of the possible outcomes for the chain helix, then, while assuming a generic helix configuration, look at the action of the magnetic field in a simple low-energy model which exposes the main idea of the paper.  One of the potential magnetic orders in the 1D chain of magnetic moments on the (super)conducting surface is attributed to the RKKY mechanism. The interaction between the Fourier modes $\bs{S}_q$ of the spins $\bs{S}_r=\sum_{q}\bs{S}_q\, e^{\mathrm{i}qr}$ at sites $r=1,\cdots, L$ of the chain is  
\begin{equation}
H_{\mathrm{RKKY}}:=
-J\sum_{q}\sum_{a,b=1}^3S_{q;a}\,\chi_{q;a,b}\,S_{-q;b},
\end{equation}
where $J>0$ is a coupling constant. In this scenario $\chi_{q;a,b}$ is the static magnetic susceptibility tensor of the (super)conducting electrons. There are differences between a 1D (which should be considered as the limit of anisotropic interactions) and 2D RKKY interaction: due to perfect nesting in 1D, the spin helix features a peak at twice the Fermi momentum $q=2k_{\mathrm{F}}$ due to resonant scattering between the Fermi points for interactions mediated by 1D normal~\cite{Braunecker09,Braunecker10,Meng13,Scheller13} or superconducting electrons~\cite{Klinovaja13}, while in the more realistic 2D case the interaction becomes short range and local ferromagnetic physics takes over \cite{Kim14}. In this latter case, intrinsic spin-orbit interaction can restore the helix. If the electronic structure is SU(2) spin-rotation symmetric, $\chi_{q;ab}$ is proportional to $\delta_{ab}$. As a consequence, the magnetic moments order in a helix structure $S_{2k_{\mathrm{F}};1}=\pm \mathrm{i}S_{2k_{\mathrm{F}};2}\neq 0$. Due to the SU(2) invariance, any rotation of the helix in spin space is a degenerate symmetry breaking state. An arbitrarily small external magnetic field $\bs{B}$ breaks the SU(2) invariance and will pin the plane in which the helix rotates to be the plane normal to $\bs{B}$.

This has to be contrasted to the situation where the magnetic chain is located on the surface of a superconductor (in the 1-2 plane, say). In this case the electrons mediating the (1D or 2D) RKKY interaction are subject to Rashba SOC due to the inversion-symmetry breaking of the surface. Along the 1D chain (the 1-direction, say), the Rashba SOC breaks the SU(2) spin-rotation symmetry down to U(1) spin-rotation symmetry in the 1-3 plane. 
This reduced symmetry is reflected by the spin susceptibility $\chi_{q;ab}$ being not diagonal anymore.
Rather, it acquires a nonvanishing off-diagonal component $\chi_{q;13}$ and the diagonal components need not all be equal, but $\chi_{q;11}=\chi_{q;33}\neq\chi_{q;22}$~\cite{Takimoto09,Santos10,Kim14}. An analysis of this for the simple but less realistic 1D RKKY case is presented in the Supplemental Material~\cite{Supp}. In particular, due to the different spin-polarizations of the Rashba-split Fermi points, the susceptibility peaks appear at different momenta in the components $\chi_{q;11}=\chi_{q;33}$ and $\chi_{q;22}$. 
As a consequence, the spin helix preferentially forms in the 1-3 plane, i.e., $S_{q_0;1}=\pm \mathrm{i}S_{q_0;3}\neq0$ for some $q_0$, while $S_{q;2}=0$, $\forall q$.
Further, the helix remains pinned to the 1-3 plane even in the presence of an external magnetic field $\bs{B}$ in the 1-2 plane, as long as $|\bs{B}|$ is smaller than some critical field $B_{\mathrm{c}}$. 
A simple estimate yields that $B_c$ is given by the SOC-induced energy splitting of the the electronic bands that mediate the RKKY interaction at the Fermi level~\cite{Supp}. If $|\bs{B}|$ is much larger than $B_{\mathrm{c}}$, the helix forms in the plane normal to $\bs{B}$ (on top of an overall homogeneous magnetization in the direction of $\bs{B}$).

In the remainder of this paper we will study the magnetically ordered chain with a magnetic field weaker than $B_{\mathrm{c}}$ in the 1-2 plane, so that the orientation of the helix is pinned to the 1-3 plane irrespective of the orientation of the external $\bs{B}$. For simplicity, we shall \textit{assume} that the magnetic helix is fully rigid and independent of $\bs{B}$. For our purposes below, this is true as long as the energy scale $B_{\mathrm{c}}$ is much larger than the $p$-wave gap induced in the chain \cite{Nadj-Perge13}. While this assumption is sufficient to demonstrate the key ideas of this proposal, a more complete analysis would have to include the influence of the magnetic field on the helical structure and the role of disorder. For example, a recent study indicates that strong disorder favors a ferromagnetic over the helical magnetic order~\cite{Kim14}. We note, however, that the effects we exploit can also occur without the helical order, if instead a suitable combination of ferromagnetic order and Rashba SOC is realized; see the Supplemental Material~\cite{Supp} for details.

\begin{figure}[t]
\centering
\includegraphics[width=.48\textwidth]{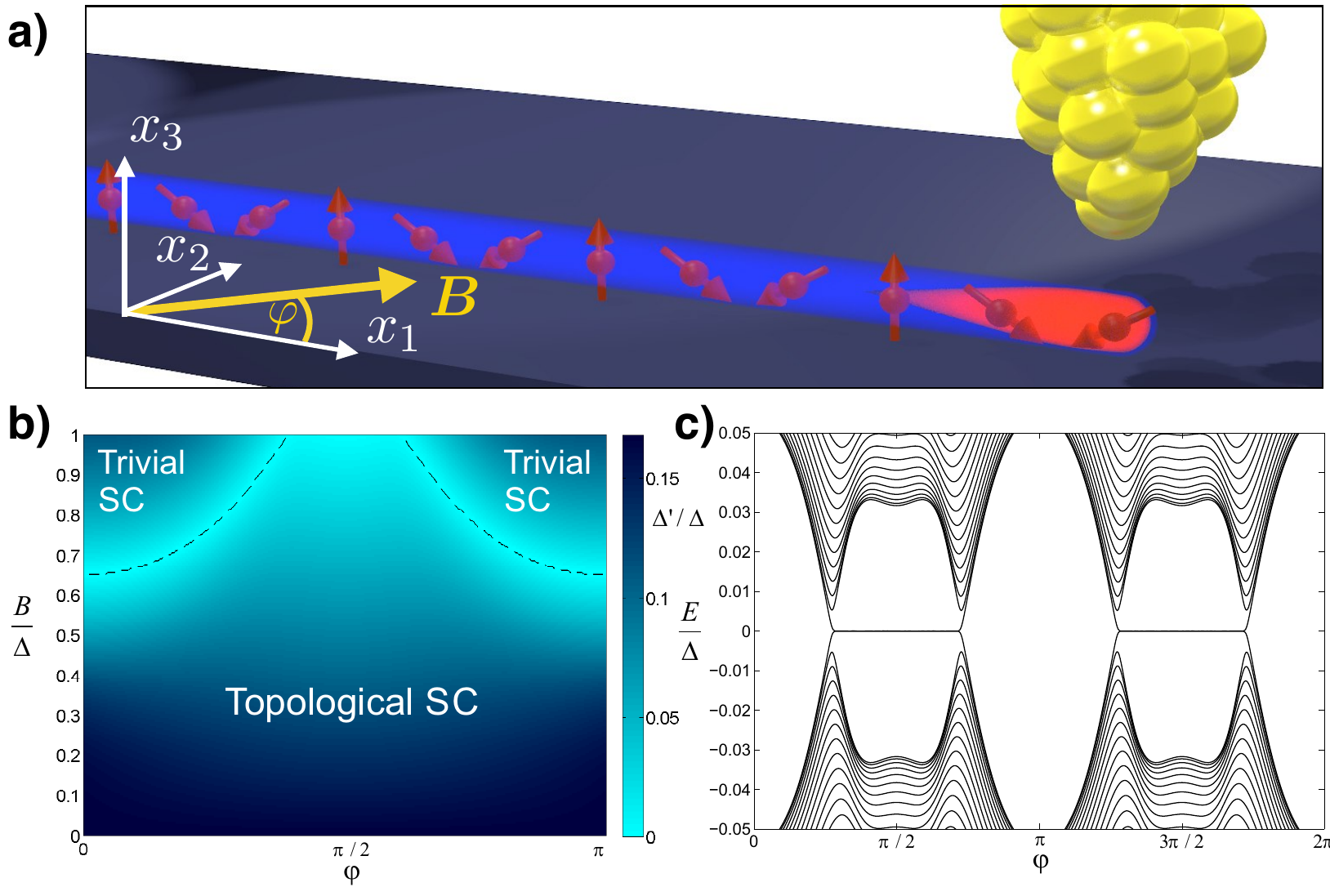}
\caption{
a) The Shiba states (blue) of a chain of magnetically ordered adatoms (red) on a thin film superconductor are a one-dimensional topological superconductor that features a Majorana end state (orange). An STM tip (yellow) can be used to detect it as a zero-bias anomaly of the tunneling current.
b) Phase diagram (bulk gap $\Delta^\prime$) of Hamiltonian~\eqref{eq: straight Wire Ham main} for 
the parameter values $\mu/\Delta=1.5$, $M/\Delta=3.4$, $\alpha/\Delta=0.75$, $t/\Delta=1$, and $\theta=\pi/6$ as a function of the angle $\varphi$ of $\bs{B}$ with the 1-direction. At constant $|\bs{B}|$, topological phase transitions are possible as a function of $\varphi$.
c) Spectrum of the finite chain of $L=600$ sites as a function of $\varphi$ for $B/\Delta=0.8$ and otherwise the same parameters. The straight lines at zero energy are Majorana zero energy states bound to the end of the chain.   
}
\label{fig: straight phases}
\end{figure}

\textit{Superconducting phases of a straight chain} --- We now want to study the effect of an external field $\bs{B}$ on the 1D chain of helical adatoms of length $L$. To that end, we consider the tight-binding Hamiltonian:  
\begin{equation}
\begin{split}
H=&\,\sum_{n=1}^L
\left\{c^\dagger_n (t+\mathrm{i}\alpha\,\sigma_2)c^{\ }_{n+1}
+\Delta c^\dagger_{n,\uparrow}c^\dagger_{n,\downarrow}
+\mathrm{h.c.}
\right\}
\\
&\,+
\sum_{n=1}^L
c^\dagger_n \left[(\bs{B}+\bs{M}_n)\cdot\bs{\sigma}-\mu\right]\,c^{\ }_{n},
\end{split}
\label{eq: straight Wire Ham main}
\end{equation}
where $c^\dagger_n=(c^\dagger_{n,\uparrow},c^\dagger_{n,\downarrow})$ and $c^\dagger_{n,s}$ creates an electron of spin $s=\uparrow, \downarrow$ on site $n=1,\cdots, L$ and the lattice spacing is unity. Here, $t$ is the nearest-neighbor hopping integral and $\Delta$ is the superconducting pairing potential. More realistic Hamiltonians for the chain involve obtaining hybridizations for the Shiba states \cite{Pientka13a,Pientka13b}. Our Hamiltonian above is, however, simple enough to illustrate the point and also quantitatively accurate for chains larger than the underlying superconductor's coherence length. The magnetic moment of the helical order (assumed to lie in the 1-3 plane) has the spatial dependence
$ \bs{M}_n=M[\cos\,(n\, \theta+\theta_0),0,\pm\sin\,(n\, \theta+\theta_0)]^{\mathsf{T}}$,
where $M$ is the overall amplitude, $\theta$ is the tilt between adjacent moments and $\pm$ stands for the two possible helicities. The overall phase $\theta_0$ has no significant effect on the spectrum of Hamiltonian~\eqref{eq: straight Wire Ham main} if the tilt angle $\theta$ is sufficiently small~\cite{Supp}.

If $|4\tilde{t}^2+\mu^2+\Delta^2-M^2|<4\mu\tilde{t}$, with $\tilde{t}:=t\cos\theta/2+\alpha\sin\theta/2$, and the external magnetic field $\bs{B}$ vanishes, the ground state of Hamiltonian~\eqref{eq: straight Wire Ham main} is a topological 1D superconductor with MBS at its end. 
We now propose to use magnetic fields in the 1- or 2-direction to drive the system into a topologically trivial state without MBS at its end. 
Upon increasing $B_2$ (with $B_1=0$), the system either enters a gapless phase at the critical field strength $B_{2,\mathrm{c}}=\Delta$ (if $M<2\tilde{t}+\mu$) or enters a gapped but topologically trivial phase at $B_{2,\mathrm{c}}=\sqrt{\Delta^2+(2\tilde{t}+\mu)^2-M^2}$ (if $M>2\tilde{t}+\mu$). 
Similarly, yet by a very different mechanism, increasing $B_1$ (with $B_2=0$) can trigger the transition into a topologically trivial gapped phase at some critical field strength $B_{1,\mathrm{c}}$. 
Unlike with $B_{2,\mathrm{c}}$, a closed analytical expression for $B_{1,\mathrm{c}}$ can only be obtained in the limit of small $\Delta$ and $B_{1}$ (see Supplementary Material~\cite{Supp}).
Crucially, the critical field strengths at which the transition occurs is generically different for fields in the 1- and 2-direction, and for appropriate choices of parameters $B_{1,\mathrm{c}}<\Delta$~\cite{Supp}. For example, if $B_{1,\mathrm{c}}<B_{2,\mathrm{c}}$ and the magnitude $B$ of the external field $\bs{B}$ is chosen such that $B_{1,\mathrm{c}}<B<B_{2,\mathrm{c}}$, a phase transition between the topological and trivial superconducting state of the chain can be crossed by \textit{rotating} $\bs{B}$ in the 1-2-plane across a critical angle $\varphi_{0}(B)$, keeping its magnitude $B$ fixed [see Fig.~\ref{fig: straight phases}(c)]. This topological phase transition as a function of the orientation of $\bs{B}$ in the 1-2-plane is the key property of the Shiba chain that we will exploit in the remainder of this work. 
It should be noted that a certain amount of fine-tuning of the model parameters is needed to simultaneously satisfy (i) $B_{1,\mathrm{c}}<B<B_{\mathrm{c}}$ to pin the magnetic structure and (ii) to maintain a sufficiently large gap for $\bs{B}$ in the 1- and 2-directions in order to protect the coherent manipulations considered below. 
For the parameters chosen in Fig.~\ref{fig: straight phases}(c), $B/\Delta=0.8<1.0\simeq B_{\mathrm{c}}/\Delta$.
If the Rashba SOC is larger, this inequality can be made more strong, e.g. for $\alpha/\Delta=5.0$, $M/\Delta=4.8$, and $\mu/\Delta=1.0$, a phase diagram similar to Fig.~\ref{fig: straight phases}(c) is obtained while $B/\Delta=0.8\ll9.0\simeq B_{\mathrm{c}}/\Delta$.

\begin{figure}[t]
\centering
\includegraphics[width=.46\textwidth]{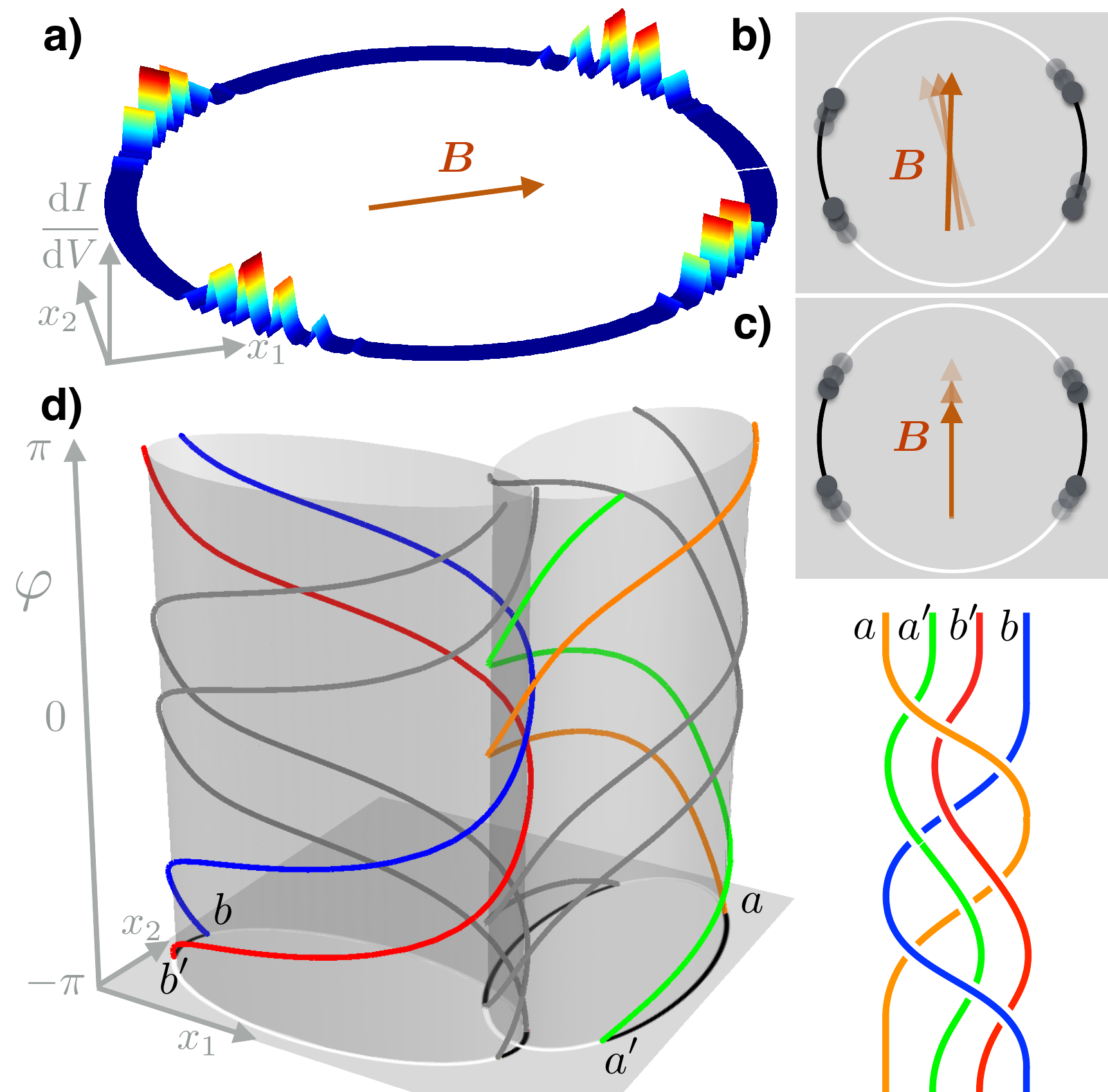}
\caption{
a) For appropriate external magnetic field strength $|\bs{B}|$, the circular magnetic chain ($L=600$ in this example) on a thin-film superconductor hosts four MBS that can be measured as zero-bias anomalies with an STM tip.
b) Changing the orientation of $\bs{B}$ moves the MBS along the circle. (Black are trivial, white are topological sectors of the chain.)
c) Changing the magnitude of $\bs{B}$ controls their distance and allows to create and fuse them pairwise.
d) Two overlapping elliptic chains can be used as a $\sigma_x$ gate that is operated by a $2\pi$ rotation of the angle $\varphi$ of $\bs{B}$ in the 1-2 plane. Shown are the world lines of the MBS during the braid. Observe that the orange MBS braids around the red, green, and blue MBS, while the green MBS braids around the blue and orange MBS only.
}
\label{fig: necklace}
\end{figure}

\textit{The necklace} --- 
In order to move Majorana bound states in a controlled way, they have to appear at some \textit{mobile} domain wall between a trivial and a topological superconducting state. Given the dependence of the topological phase transition on the orientation of the external magnetic field, such domain walls can occur between chain segments with different relative orientations to $\bs{B}$.
One can join up many such segments into a bent chain or a circle. 
For the latter, the angle $\varphi$ between the homogeneous external field $\bs{B}$ and (the tangential to) the circular chain becomes position dependent as $\varphi\to\varphi_n=2\pi n/L$. If $L\gg \theta^{-1}$, the local magnetic structure of the chain will not be affected by the bending. Locally, the helical magnetic moment is taken to lie in the plane spanned by the tangential to the circle and the 3-direction. We choose the length $L$ such that no frustration or magnetic domain wall are induced due to the periodic boundary conditions, i.e., $L\theta/(2\pi)\in\mathbb{Z}$.  
Then, if $B_{1,\mathrm{c}}<B<B_{2,\mathrm{c}}$, two segments of the circle with angles 
$\varphi\in[-\varphi_{0}(B),\varphi_{0}(B)]$ and $\varphi\in[\pi-\varphi_{0}(B),\pi+\varphi_{0}(B)]$ are trivial while the two segments separating them are topological one-dimensional superconductors [see Fig.~\ref{fig: necklace}(a)]. Each of the four domain walls between these segments host a single MBS.
For any finite $L$, the finite-size splitting of the MBS due to their hybridization is exponentially small in $L/\xi$, with $\xi$ the coherence length. Note that the energy separation of the MBS from excited states also scales to zero in the thermodynamic limit $L\to \infty$, because the circle and with it also the shape of the  domain wall are self-similar for different $L$.  The situation is similar to that of a very soft boundary between topological and nontopological phases. However, the excitation energy only scales to zero as $\xi/L$ or slower~\cite{Supp}. For a finite circle, this weak scaling leaves several orders of magnitude in energy between the MBS splitting and the lowest excitation energy.  For example, with the parameters used in Fig.~\ref{fig: straight phases}(b) and $L=600$, the MBS splitting is $10^{-5}\,\Delta$ while the energy of the first excited state is $0.035\,\Delta$. On the intermediate energy (or time) scales, we now can perform dynamical operations on the MBS. For example, the four MBS on the circle can be moved along the circle by simply rotating the magnetic field, while maintaining their relative positions~[see Fig.~\ref{fig: necklace}(b)].

\begin{figure}[t]
\centering
\includegraphics[width=.46\textwidth]{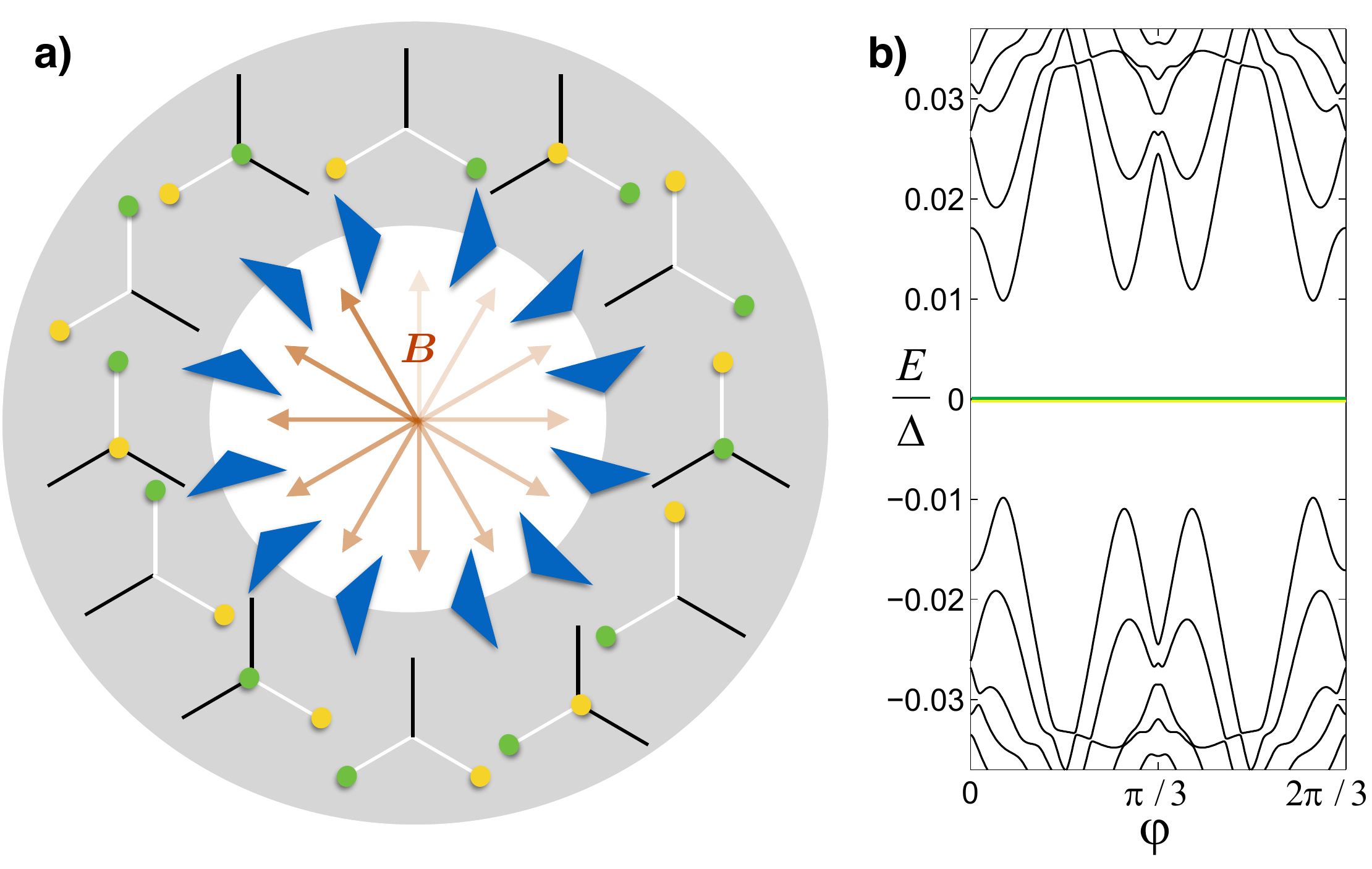}
\caption{
a) A trijunction of chains in an inplane magnetic field $\bs{B}$ supports two MBS and can be used to braid two MBS via a $2\pi$ rotation of $\bs{B}$.
b) Spectrum of the trijunction as a function of the orientation $\varphi$ of $\bs{B}$. The spectrum is $2\pi/3$-periodic. Each chain of length $L=180$ is governed by Hamiltonian~\eqref{eq: straight Wire Ham main} with the same parameters as in Fig.~\ref{fig: straight phases}(b).
}
\label{fig: trijunction}
\end{figure}

\textit{The $\sigma_x$ gate} --- 
The control over magnitude and angle of the external magnetic field is sufficient to create and annihilate qubits of MBS and to perform gate operations on the qubits, as we now show. Two distant MBS $a$ and $a'$ form a nonlocal fermionic two-level system with occupation eigenstates $|0\rangle_a$ and $|1\rangle_a$ for zero and one fermion, respectively. All topologically protected adiabatic operations, such as the rotation of the external field, do not change the fermion parity of the superconducting chain. Hence, qubit operations have to be performed on a Hilbert space of constant parity. For that reason, we consider a qubit made of two pairs $a, a'$ and $b, b'$ of MBS. 
Specifically, we consider a setup of two intersecting ellipses as sketched in Fig.~\ref{fig: necklace}~(d), where the MBS pair $a,a'$ is located on one ellipse, while the pair $b,b'$ is on the other.
Figure~\ref{fig: necklace}(d) shows the world lines of the MBS under a $2\pi$ rotation of the external field $\bs{B}$. This process
braids the MBS $a$ with $a'$, $b$, and $b'$, while it braids $a'$ with $a$ and $b$, but not with $b'$. As a result, the qubit formed from $a$, $a'$, $b$, and $b'$ changes as
\begin{equation}
|\bar{1}\rangle_{ab}\to|\bar{0}\rangle_{ab},
\qquad
|\bar{0}\rangle_{ab}\to|\bar{1}\rangle_{ab},
\label{eq: sigma x gate operation}
\end{equation}
where we denoted  the qubit eigenstates as $|\bar{0}\rangle_{ab}:=|0\rangle_a\otimes |0\rangle_b$ and $|\bar{1}\rangle_{ab}:=|1\rangle_a\otimes |1\rangle_b$.
This constitutes a $\sigma_x$ gate operation. Each ellipse in Fig.~\ref{fig: necklace}(d) hosts a second pair of MBS, whose world lines are indicated in grey. These MBS do not interfere with the qubit operation on $a,a'$ and $b,b'$. Rather, they form a second qubit on which the same operation is performed.
Up to a global phase, the transformation~\eqref{eq: sigma x gate operation} depends only on the topology of the worldlines of the MBS and not on any details of the adiabatic evolution. We have confirmed Eq.~\eqref{eq: sigma x gate operation} via an explicit calculation of the adiabatic evolution of the many-body states under the field rotation using a simplified model (see Supplemental Material~\cite{Supp}, Sec.~C~3).

\textit{The braiding operation} --- 
We now focus on a qubit composed of a single pair of MBS $a$ and $a'$ and want to perform a braiding of the two MBS. This operation conserves the parity of the qubit. We use a trijunction of straight chains to demonstrate the operation (see Fig.~\ref{fig: trijunction}). When choosing $\varphi_0(B)\in(\pi/6,\pi/3)$, the trijunction hosts two MBS with one or two topological segments for every inplane orientation of $\bs{B}$. As sketched in Fig.~\ref{fig: trijunction} (a), a rotation of $\bs{B}$ by $2\pi$ braids the two MBS around one another. In this process, the states of the qubit transform as $|0\rangle_a\to|0\rangle_a$ and $|1\rangle_a\to-|1\rangle_a$, modulo a nonuniversal overall U(1) phase factor multiplying both basis states. Figure~\ref{fig: trijunction} (b) shows the spectrum of the trijunction during the braid. It should be noted that the gap that protects the MBS collapses for certain orientations of $\bs{B}$ as the number of sites in each segment is taken to infinity. However, the finite size gap can be several orders of magnitude larger than the splitting between MBS. (The gaps are $0.01\Delta$ and $10^{-6}\Delta$, respectively, for the parameter values studied in Fig.~\ref{fig: trijunction}(b).)

\textit{Measurement and experimental implementation} --- 
We propose local experimental probes, namely an STM and a scanning single-electron transistor (SET) to demonstrate that the MBS (i) exist, (ii) can be moved and (iii) can be braided in the proposed necklace setup.
To show (i), STM probes zero bias anomalies (ZBA) at four positions along the circle. While open wires might host nontopological end states that can be confused with the MBS, ZBAs in the necklace directly evidence the magnetic field induced topological phase transition.
To show (ii), STM measurements at different orientations of in-plane magnetic field, will find the MBS at different locations along the chain. As long as $B<B_{1,\mathrm{c}}$, STM measurements will show no ZBA at any location along the necklace. For $B_{2,\mathrm{c}}>B>B_{1,\mathrm{c}}$, four ZBAs will appear in every rotation period of $\bs{B}$.
As these measurements are sensitive to the density of states of the MBS, but not their quantum state, they do not suffer from decoherence processes.
In contrast, to show (iii), the quantum state of the MBS has to be prepared before the gate operation and measured thereafter. We propose the following protocol using the setup of Fig.~\ref{fig: necklace}. First, $B$ is increased adiabatically beyond $B_{1,\mathrm{c}}$, at which point pairs of MBS are created. Due to the finite overlap of the MBS, each pair's eigenstates $|0\rangle$ and $|1\rangle$ are split in energy and each is prepared in the lower state $|0\rangle$. Second, for appropriate $B_{2,\mathrm{c}}>B>B_{1,\mathrm{c}}$, the $\sigma_x$ gate operation is implemented by a $2\pi$ rotation of $\bs{B}$. This changes the state of each of the four pairs of MBS to $|1\rangle$, taking two Cooper pairs from the condensate. Third, the MBS are fused pairwise by reducing the magnetic field $B\to B_{1,\mathrm{c}}$. Integration of a scanning SET within an STM step-up will be required to probe the local charge density and sense the presence (or absence) of remnant charge and state of the Majorana qubits after manipulation. It is important not to use the STM tunneling at all during such manipulation, as tunneling of quasiparticles from an STM tip will cause decoherence. A key advantage of our proposed approach is that metal contacts or gates (that are sources of unpaired poisoning quasiparticles) are not required for manipulation of MBS.   

We close with some crude estimates of the energy and time scales relevant to the experimental realization of our braiding proposal. The detection of MBS through the STM measurement can be made in a slowly rotating magnetic field and hence is not subject to the more strict requirements necessary for braiding. Quasiparticle poisoning can be as hazardous to this as to other platforms that realize MBS~\cite{Fu09,Heck11, Hassler11,Rainis12,Liu13}. It reduces the height of the ZBA and is believed to be the main source of decoherence during a braiding operation. Typically, the thin-film superconducting gap $\Delta\sim 1\, \text{meV}$, so that the gap protecting the MBS in the straight chain or necklace $\Delta_{\text{necklace}}\sim 0.05\, \text{meV}$. We can estimate thermal quasiparticles to appear at a rate $\omega_{\mathrm{qp}}\sim0.001\,\text{meV}$ (which is much larger than the finite size MBS splitting). The challenge is to perform the field rotation of the field $B\sim 0.5\,\mathrm{T}$ at a frequency $\omega$ that is quasi adiabatic wrt.\ $\Delta_{\text{necklace}}$ but large wrt.\ $\omega_{\mathrm{qp}}$, i.e., $\omega\sim0.01\,\text{meV}\sim2\,\text{GHz}$. A setup in which current pulses running through two perpendicular superconducting wires create a rotating magnetic field can be envisioned~\cite{Supp}.

\textit{Conclusions} --- 
Utilizing the Shiba states of magnetic adatoms on the surface of a thin film superconductor, we have proposed a system
that allows for the simple detection and manipulation of MBS and in particular can be used to demonstrate their non-Abelian character.  
Our proposal takes full advantage of the high level of control that one has regarding the design, operation, and measurements using STM in this setup.

\textit{Acknowledgement} --- The authors acknowledge financial support by the ARO MURI program W911NF-12-1-0461, the NSF-DMR1104612, ONR-N00014-11-1-0635, NSF-MRSEC
NSF-DMR0819860 programs (A.Y.), by
DARPA SPAWARSYSCEN Pacific N66001-11-1-4110 (T.N., B.A.B, A.Y.), by  Packard Foundation, MURI-130-6082, Keck Foundation, NSF CAREER DMR- 095242, ONR - N00014-11-1-0635 and NSF-MRSEC DMR-0819860 (B.A.B.) and by the Swiss National Science Foundation (J.L.).

\addtocontents{toc}{\SkipTocEntry}

\appendix
\begin{widetext}
\newpage
\begin{center}
\large
\textbf{
Supplemental Material for ``Majorana zero modes on a necklace''}
\normalsize
\end{center}

\tableofcontents

\section{Magnetic structure of a straight chain}
\label{sec: magnetism}

We consider the classical magnetic order that emerges due to $1$D RKKY interaction mediated by electrons that are subject to an extra Rashba spin-orbit coupling. The Rashba spin-orbit coupling induces Dschaloschinskii-Moriya interactions between the spins, resulting in a helical magnetic order of preferred orientation. The effect of an externally applied Zeeman field is studied. If large enough, the field can change the direction in which the helix forms. We stress that these calculations are valid only for pure $1$D RKKY interaction, which is not the realistic situation in a $2$D sample. However, it is simple enough to be solved analytically, and gives a hint of the effects of an external magnetic field action. 

\subsection{RKKY interactions in presence of Rashba spin-orbit coupling}

Consider electrons of  effective mass $m$ that are confined in a one-dimensional channel and governed by the second-quantized Hamiltonian
\begin{equation}
H_0:=
\sum_k c^\dagger_k\left(\frac{k^2}{2m}-\mu+\bs{g}_k\cdot\bs{\sigma}\right)c^{\ }_k,
\label{eq: continuum noninteracting H}
\end{equation}
where $c_k=(c_{k,\uparrow},c_{k,\downarrow})^{\mathsf{T}}$, and $c^\dagger_{k,s}$ creates an electron of momentum $k$ and spin $s=\uparrow, \downarrow$. The vector $\bs{g}_k=-\bs{g}_{-k}$ parametrizes the spin-orbit coupling in the system and $\bs{\sigma}=(\sigma_1,\sigma_2,\sigma_3)$ is the vector of the three Pauli matrices acting in spin space. 
The spectrum of $H_0$ consists of two spin-orbit split branches and is given by
\begin{equation}
\xi_{k,\lambda}:=\frac{k^2}{2m}-\mu+\lambda|\bs{g}_k|,
\qquad \lambda=\pm.
\end{equation}

The conduction electrons couple via the Hund's coupling of strength $\beta$
\begin{equation}
H_{\mathrm{H}}:=\beta
\sum_{k,q}\bs{S}_q\cdot\left(c^\dagger_{k-q}\bs{\sigma}c^{\ }_{k}\right)
\end{equation}
to localized magnetic moments  $\bs{S}_r=\sum_{q}\bs{S}_q\, e^{\mathrm{i}qr}$ on sites $r$ that are arranged in a one-dimensional chain.
The conduction electrons induce an effective RKKY interaction between the localized spins  
\begin{equation}
H_{\mathrm{RKKY}}:=
-\frac{2\beta^2}{A^2}\sum_{q}\sum_{a,b=1}^3S_{-q;a}\,\chi_{q;a,b}\,S_{q;b},
\label{eq: classical Hamiltonian operator in momentum space}
\end{equation}
where $A$ is a constant with the dimension of area that characterizes the extend of the electron wavefunction perpendicular to the chain and $\chi_{q;a,b}$, $a,b=1, 2, 3$, is the static magnetic susceptibility tensor of the electrons. 

We now assume a geometry in which the conduction electrons originate from a two-dimensional surface on which the magnetic spins are positioned, that has its normal in the 3-direction. The Rashba spin-orbit coupling due to the inversion symmetry breaking by the surface takes the form $\bs{g}_{\bs{k}}\propto\bs{e}_3\wedge \bs{k}$, with $\bs{e}_3$ the unit vector in 3-direction. If we further assume that the chain of magnetic moments stretches along the 1-direction on the surface, so that $k\equiv k_1$, the spin-orbit coupling term in Hamiltonian~\eqref{eq: continuum noninteracting H} takes the from $\bs{g}_k=(0,\alpha\,k,0)^{\mathsf{T}}$. 
The electron's spin susceptibility is then 
\begin{equation}
\chi^{\ }_{q;ab}
\propto
\int\mathrm{d}k\int\mathrm{d}\omega
\mathrm{tr}\left[\sigma_aG_{k+q,\omega}\sigma_bG_{k,\omega}\right],
\end{equation}
where $G_{k+q,\omega}$ is the Green function corresponding to Hamiltonian~\eqref{eq: continuum noninteracting H}. Hamiltonian~\eqref{eq: continuum noninteracting H} has a spin rotation symmetry if $\bs{g}_{\bs{k}}\propto\bs{e}_3\wedge \bs{k}$ for rotations around the 2-direction in spin space $\mathcal{R}_2(\varphi)$ by an arbitrary angle $\varphi\in[0,2\pi)$, i.e., $\mathcal{R}_2(\varphi)H_0\mathcal{R}^{-1}_2(\varphi)=H_0$,  and so has the Green function.
As a consequence, for $b=1,3$,
\begin{equation}
\begin{split}
\chi^{\ }_{q;2b}
&\propto
\int\mathrm{d}k\int\mathrm{d}\omega
\mathrm{tr}\left[\sigma_a\mathcal{R}_2(\pi)G_{k+q,\omega}\mathcal{R}^{-1}_2(\pi)
\sigma_b\mathcal{R}_2(\pi)G_{k,\omega}\mathcal{R}^{-1}_2(\pi)\right]
\\
&
=
\int\mathrm{d}k\int\mathrm{d}\omega
\mathrm{tr}\left[\sigma_2G_{k+q,\omega}\mathcal{R}^{-1}_2(\pi)
\sigma_b\mathcal{R}_2(\pi)G_{k,\omega}\right]
\\
&
=
-\int\mathrm{d}k\int\mathrm{d}\omega
\mathrm{tr}\left[\sigma_2G_{k+q,\omega}
\sigma_bG_{k,\omega}\right]
\\
&
\propto
-\chi^{\ }_{q;2b},
\end{split}
\label{eq: vanishing terms in the spin susceptibility}
\end{equation}
where we used $\mathcal{R}^{-1}_2(\pi)\sigma_b\mathcal{R}_2(\pi)=-\sigma_b$ for $b=1,3$. Thus, the components $\chi^{\ }_{q;21}$, $\chi^{\ }_{q;23}$, $\chi^{\ }_{q;12}$, and $\chi^{\ }_{q;32}$ vanish for reasons of symmetry. 
The remaining components of the electron's spin susceptibility can be computed from
\begin{equation}
\begin{split}
\hat{\chi}^{\ }_{q}=&
\frac{1}{N}
\sum_{k}
\sum_{\lambda,\lambda'=\pm}
\begin{pmatrix}
1-s_{k(k+q)}\lambda\lambda'&0&\mathrm{i}(\lambda s_k-\lambda' s_{k+q})\\
0&1+s_{k(k+q)}\lambda\lambda'&0\\
-\mathrm{i}(\lambda s_k-\lambda' s_{k+q})&0&1-s_{k(k+q)}\lambda\lambda'
\end{pmatrix}
\frac{
f_{\mathrm{FD}}(\xi_{k,\lambda})-f_{\mathrm{FD}}(\xi_{k+q,\lambda'})     }
     {
\xi_{k+q,\lambda'}-\xi_{k,\lambda}     
     },
\end{split}     
\label{dynsuscept zero T 1D notmal state}
\end{equation}
where $N$ is the number of lattice sites, $f_{\mathrm{FD}}$ is the Fermi-Dirac distribution, and $s_k:=\mathrm{sign}\,k$. Two observations are in order. (i) In absence of Rashba spin-orbit interaction ($\xi_{k,\lambda}=\xi_{k,\lambda'}$), the susceptibility tensor takes the SU(2) invariant form $\hat{\chi}_q=\openone\,\chi_q$. In this case $\chi_q$ features a divergence at $q=2 k_{\mathrm{F}}:=2\sqrt{2m\mu}$ which leads to a helical magnetic ordering with that wavevector. The magnetic order spontaneously breaks both the continuous SU(2) rotation symmetry in spin space as well as the discrete translational symmetry. 
(ii) With finite Rashba spin-orbit coupling, the diagonal entries of the 1 and 3 components remain equal (degenerate) and are coupled by off-diagonal matrix elements of definite helicity.

\begin{figure}[t]
\centering
\includegraphics[width=.86\textwidth]{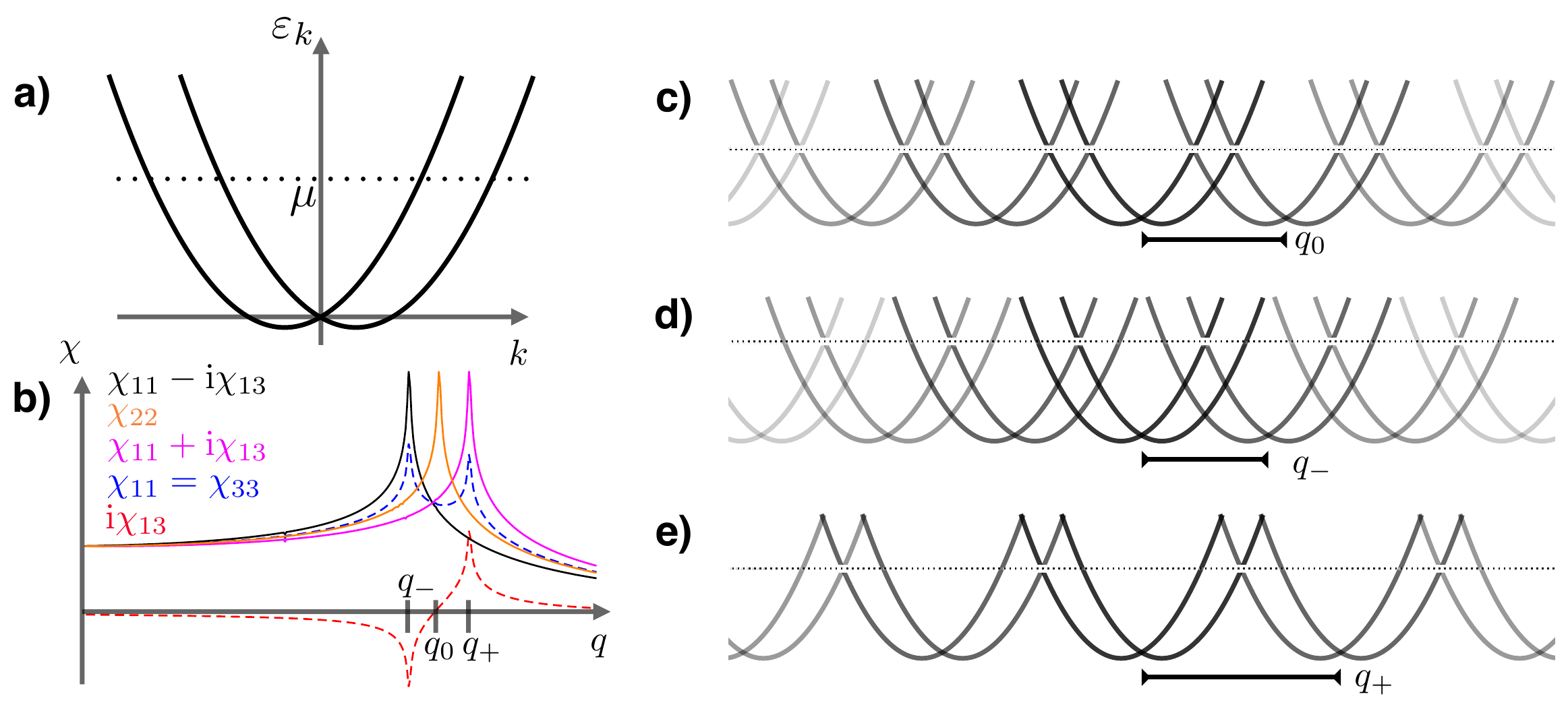}
\caption{
a) Schematic band structure of the Rashba Hamiltonian~\eqref{eq: continuum noninteracting H}.
b) Various components of the susceptibility tensor for the non-superconducting state with Rashba spin-orbit coupling. The single peak or pair of peaks appear in the components $\chi_{22}$ and $\chi_{11}=\chi_{33}$, respectively, where the splitting between the peaks is determined by the spin-orbit coupling length. 
c)-e) Band hybridization caused by the three magnetic instabilities (spin-density wave and two helical spin states with opposite helicity) at the ordering wave vectors $q_0$ and $q_\pm$.
}
\label{fig: susceptiblity}
\end{figure}

Numerical evaluation of the nonvanishing matrix elements of $\hat{\chi}_q$ shows that $\chi_{q;11}=\chi_{q;33}$ develops a double peak structure with one peak each at 
\begin{equation}
q_{\pm}:=2\sqrt{2m\mu+m^2\alpha^2}\pm 2m\alpha,
\label{eq: peak in chi 11}
\end{equation} 
while $\chi_{q;22}$ retains a single peak at $q_0:=2\sqrt{2m\mu+m^2\alpha^2}$ [Consult Fig.~\ref{fig: susceptiblity} (c)-(e) for the physical significance of these momenta.]. 
The purely imaginary off-diagonal terms $\chi_{q;13}=-\chi_{q;31}$ are peaked at  $q_{\pm}$ as well and change sign between the peaks
$\chi_{q_+;13}=-\chi_{q_-;13}$ [see Fig.~\ref{fig: susceptiblity} (b)].

The dominant magnetic instability is determined by the peak structure of the magnetic susceptibility [see Fig.~\ref{fig: susceptiblity} (b)]. At first sight, it seems that the peak in $\chi_{q,22}$ is far exceeding all other peaks and one would suspect a spin-density wave order with $q_0$ would form, hybridizing the bands as shown in Fig.~\ref{fig: susceptiblity} (c). However, the susceptibilities for the helical orders in the 1-3-plane are in fact $\chi_{q,11}\pm\mathrm{i}\chi_{q,13}$ and we observe from Fig.~\ref{fig: susceptiblity} (b) that they have degenerate peak heights. [Precise definitions of these magnetic orders are given in the next section.]

\subsection{The classical magnetic order}

We now want to find the classical ground state of Hamiltonian~\eqref{eq: classical Hamiltonian operator in momentum space}. To be able to make analytical progress, we make simplifying assumptions about the form of the susceptibility tensor. For one, Eq.~\eqref{eq: vanishing terms in the spin susceptibility} enforces by symmetry zeros in the susceptibility tensor that render it block-diagonal, decoupling the 2-components of the magnetic moments from the 1- and 3-components of the magnetic moments, with the latter mutually coupled. 
We shall approximate the various peaks in the susceptibility by a delta function in momentum space.

We want to solve for the spin configuration that minimizes the energy of Hamiltonian~\eqref{eq: classical Hamiltonian operator in momentum space} if the local constraint
\begin{equation}
\bs{S}^2_r=1,\quad\forall r=1,\cdots,L,
\label{eq: constraint}
\end{equation}
is imposed on each site $r$ of the chain of length $L$
for the two cases
\begin{equation}
\begin{split}
\text{(A)}\qquad
&
\chi_{q,22}=\frac{\bar{\chi}}{2}(\delta_{q-q_0}+\delta_{q+q_0}),
\\
\text{(B)}\qquad
&
\chi_{q,11}=\chi_{q,33}=\frac{\bar{\chi}_{11}}{2}(\delta_{q-q_+}+\delta_{q+q_+}),
\\
&
\chi_{q,13}=\chi_{q,31}^*=-\frac{\bar{\chi}_{13}}{2}(\delta_{q-q_+}-\delta_{q+q_+}),
\end{split}
\end{equation}
with all components not listed vanishing in either case.

\subsubsection{Case (A): Antiferromagnet}

In case (A), in view of Hamiltonian~\eqref{eq: classical Hamiltonian operator in momentum space},
the task is to maximize the value of $|S_{2;q_0}|^2$ for spin configurations subject to the constraint~\eqref{eq: constraint}. The Fourier component of a spin configuration $S_{2;r}=\pm1$ for all sites $r$ is computed as
\begin{equation}
S_{2;q_0}=\frac{1}{L}\sum_r e^{-\mathrm{i} q_0 r}S_{2;r}.
\end{equation}
Consider without loss of generality the case where $S_{2;q_0}=S_{2;-q_0}$ is real, 
\begin{equation}
\frac{S_{2;q_0}+S_{2;q_0}}{2}=
\frac{1}{L}\sum_r \cos( q_0 r) S_{2;r}.
\end{equation}
To maximize the left hand side, we have to choose
\begin{equation}
S_{2;r}=\mathrm{sign}[\cos( q_0 r)].
\end{equation} 
This configuration has the Fourier component of size
\begin{equation}
S_{2;q_0}=\frac{2}{\pi}\approx 0.637
\end{equation}
and the energy is 
\begin{equation}
E_{\mathrm{af}}=-\frac{2\beta^2}{A^2}\left(\frac{2}{\pi}\right)^2\bar{\chi}
\approx-\frac{2\beta^2}{A^2} 0.405  \bar{\chi}.
\label{eq: af energy}
\end{equation}
Note that this is in particular smaller than the energy $E_{\mathrm{af}}
\approx-\frac{2\beta^2}{A^2} 0.25  \bar{\chi}$ of a spin-density wave $S_{2;r}=\cos( q_0 r)$ [which does not satisfy the local constraint~\eqref{eq: constraint}].

\subsubsection{Case (B): Helix}
To analyze case (B), we have to face the problem that the local constraint~\eqref{eq: constraint} is not conveniently written in momentum space. To circumvent this, we replace it by the global constraint 
\begin{equation}
\sum_r\bs{S}^2_r=L
\end{equation}
and check a posteriori that the solution we obtained also satisfies the local constraint. 
We thus consider the Hamiltonian
\begin{equation}
\begin{split}
H_{\mathrm{helix}}:=
&-\frac{2\beta^2}{A^2}
\left[
\bar{\chi}_{11} S_{-q_+;1}S_{q_+;1}+\bar{\chi}_{11} S_{-q_+;3}S_{q_+;3}
-\bar{\chi}_{13}(S_{-q_+;1}S_{q_+;3}-S_{-q_+;3}S_{q_+;1})
\right]
\\
&-\lambda
\sum_{q}
\bs{S}_{-q}\cdot \bs{S}_{q},
\end{split}
\label{eq: helix Hamiltonian}
\end{equation}
where $\lambda$ is a Lagrange multiplier.

Minimization yields
\begin{equation}
\begin{split}
&0=\frac{\partial H_{\mathrm{helix}}}{\partial S_{\pm q_+;1}}
=-\frac{2\beta^2}{A^2}(\bar{\chi}_{11} S_{\mp q_+;1}\pm\bar{\chi}_{13}S_{\mp q_+;3})-\lambda S_{\mp q_+;1},
\\
&0=\frac{\partial H_{\mathrm{helix}}}{\partial S_{\pm q_+;3}}
=-\frac{2\beta^2}{A^2}(\bar{\chi}_{11} S_{\mp q_+;3}\mp\bar{\chi}_{13}S_{\mp q_+;1})-\lambda S_{\mp q_+;3},
\\
&0=\frac{\partial H_{\mathrm{helix}}}{\partial \lambda}
=-\sum_{q}\bs{S}_{-q}\cdot \bs{S}_{q}.
\end{split}
\label{eq: minimization}
\end{equation}
A solution to Eq.~\eqref{eq: minimization} is given by
\begin{equation}
S_{\pm q_+;1}=\frac{1}{2},
\qquad
S_{\pm q_+;3}=\pm\frac{1}{2\mathrm{i}},
\qquad
\lambda=-\frac{2\beta^2}{A^2}(\bar{\chi}_{11}+\mathrm{i}\bar{\chi}_{13}).
\end{equation}
This is nothing but a helix in position space
\begin{equation}
\bs{S}_r=(\cos(q_+r),0,\sin(q_+r))^{\mathsf{T}},
\label{eq: helix plus}
\end{equation}
which automatically satisfies the local constraint~\eqref{eq: constraint} as well.
The energy of this minimizing solution is
\begin{equation}
E_{\mathrm{helix}}=-\frac{2\beta^2}{A^2}\frac{\bar{\chi}_{11}+\mathrm{i}\bar{\chi}_{13}}{2},
\end{equation}
which is lower than the energy~\eqref{eq: af energy}, given the observation that $\bar{\chi}_{11}+\mathrm{i}\bar{\chi}_{13}=\bar{\chi}$.

The same arguments for the wave vector $q_-$ yield the conclusion that a helix with opposite helicity
\begin{equation}
\bs{S}_r=(\cos(q_-r),0,-\sin(q_-r))^{\mathsf{T}},
\label{eq: helix minus}
\end{equation}
is an energetically degenerate state. 

We conclude that the magnetic state that minimizes the energy is given by one of the degenerate helices~\eqref{eq: helix plus} and~\eqref{eq: helix minus}.
The back-action of this helical magnetic order on the itinerant electronic states is the opening of a hybridization gap between two of the four electronic branches that cross the Fermi level [see Fig.~\ref{fig: susceptiblity} (d) and (e)]. We are thus left with a single species of effectively spineless electrons. When superconducting paring is induced on them by proximity, they are bound to form a topological superconductor (in this toy unrealistic model of 1D RKKY interaction).
One might wonder how the presence of superconductivity alters the RKKY mechanism presented here. In fact, as long as the superconducting order parameter is smaller than the energy scale associated with the Rashba spin-orbit coupling ($\Delta\ll\alpha k_{\mathrm{F}}$), its effect is merely to round off the peaks in the magnetic susceptibility, leaving the qualitative results unaltered.

\subsection{The effect of a finite in-plane magnetic field}

\begin{figure}[t]
\centering
\includegraphics[width=.94\textwidth]{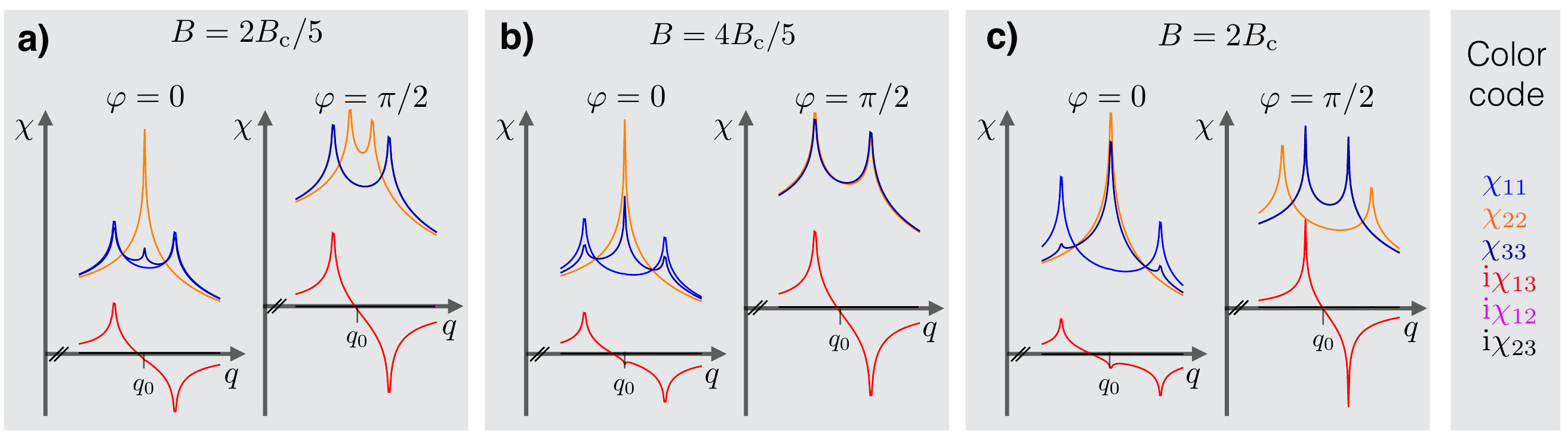}
\caption{Various components of the susceptibility tensor for the non-superconducting state with Rashba spin-orbit coupling and an external magnetic field according to Eq.~\eqref{eq: B field}. 
(a) $B=0.2$, (b) $B=0.4$, which is close to $B_{\mathrm{c}}$ defined in Eq.~\eqref{eq: critical field},
and (c) $B=1.0$. 
The curves for $\mathrm{i}\chi_{12}$ and  $\mathrm{i}\chi_{23}$ are degenerate along the horizontal axis in all panels. The curves for $\chi_{11}$ and  $\chi_{33}$ are degenerate for $\varphi=\pi/2$ in all panels.
The momentum $q_0$ is defined below Eq.~\eqref{eq: peak in chi 11}.
}
\label{fig: susceptiblity in field}
\end{figure}

In anticipation of the crucial role that an externally applied homogeneous magnetic field will play in our proposal, we want to study whether such a field will change the nature of the magnetic order. In particular, we are interested in applying a field in the 1-2 plane. This field does not destroy the superconductivity of a thin layer in the 1-2 plane, since the orbital repairing is not effective. 
We use the parametrization
\begin{equation}
\bs{B}=B(\cos\,\varphi,\sin\,\varphi,0),
\label{eq: B field}
\end{equation}
which essentially alters the electronic structure via Zeeman coupling according to the substitution
\begin{equation}
\bs{g}_k=(B\,\cos\,\varphi,B\,\sin\,\varphi+\alpha k,0)
\end{equation}
in Hamiltonian~\eqref{eq: continuum noninteracting H}.
Our strategy is to compute the magnetic susceptibility in the presence of the magnetic field~\eqref{eq: B field} and to deduce possible changes in the magnetic order from the modifications in the peak structure. 
Such changes may indicate transitions between phases of qualitatively different magnetic order that might be induced by the external magnetic field at some critical field strength $B_{\mathrm{c}}$.

Studying the evolution of the Fermi points as a function of $\varphi$ and $B$ can already give a hint. At the magnetic field determined by the Rashba spin-orbit coupling energy scale
\begin{equation}
B_{\mathrm{c}}
=\sqrt{2 m \mu}\alpha
\label{eq: critical field}
\end{equation}
two of the four Fermi points at positive (negative) momenta become degenerate for the field orientation $\varphi=\pi/2$ ($\varphi=3\pi/2$).

\begin{figure}[t]
\centering
\includegraphics[width=.4\textwidth]{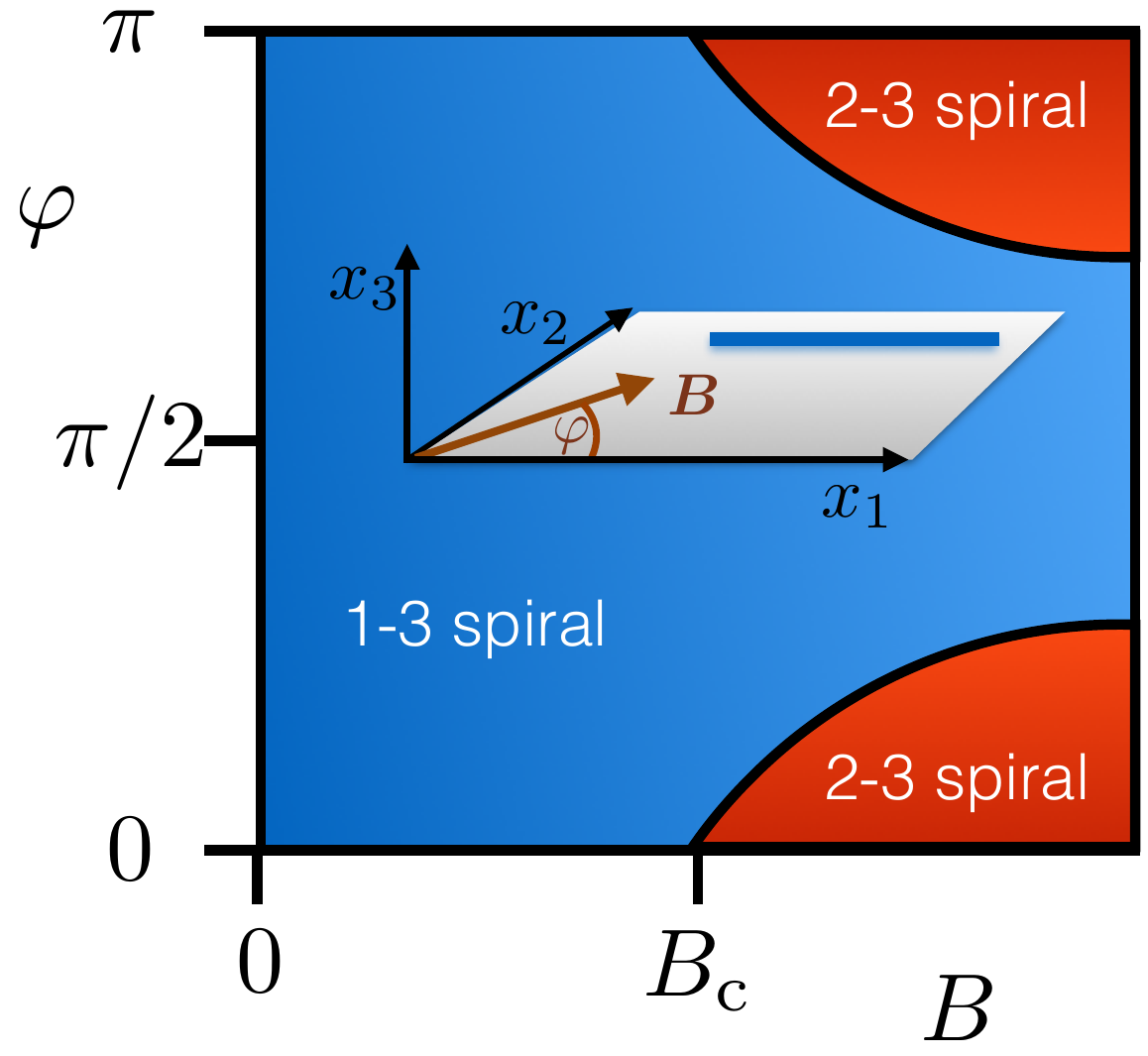}
\caption{Schematic phase diagram for the influence of a homogeneous magnetic field on the helical magnetic order. If $B\ll B_{\mathrm{c}}$ a helix in the 1-3 plane is formed independent of the field orientation $\varphi$. For large magnetic field, in contrast, we observe a tendency for forming a helix in the plane perpendicular to the magnetic field.
This phase diagram only shows the effect of the homogeneous magnetic field on the spatially varying (helical) component of the magnetization. In addition to the modification of the helical order, the homogeneous magnetic field will also induce an overall homogeneous magnetization parallel to the field, which will grow with $B$.
}
\label{fig: magnetic phase diagram}
\end{figure}

A numerical evaluation of the magnetic susceptibility is shown in in Fig.~\ref{fig: susceptiblity in field}. For simplicity, the susceptibility is only shown for the special field orientations $\varphi=0$ and $\varphi=\pi/2$, but the following discussion is also compatible with the interpolation between these cases.  
For $B<B_{\mathrm{c}}$, the changes in the magnetic susceptibility are small, suggesting that the helical order in the 1-3 plane is stable [see Fig.~\ref{fig: susceptiblity in field} (a)].
As $B$ approaches $B_{\mathrm{c}}$, at $\varphi=0\mathrm{mod}\,\pi$ an extra peak appears in $\chi_{33}$ at $q_0$ [see panel (b) from Fig.~\ref{fig: susceptiblity in field}]. This, together with the already existing peak of $\chi_{22}$ at $q_0$ poses the possibility of helical magnetic order in the 2-3 plane, that competes with the order in the 1-3 plane. At the same value of the magnetic field, but for $\varphi=\pi/2\mathrm{mod}\pi$, the single peak in $\chi_{22}$ at $2k_{\mathrm{F}}$ splits and might allow for helical order in the 1-2 or 2-3 plane. However, since the components $\chi_{12}$ and $\chi_{23}$ are much smaller than $\chi_{13}$ (and in fact very close to zero in Fig.~\ref{fig: susceptiblity in field}), the helix in the 1-3 plane is likely to remain the dominant instability at this angle of the magnetic field. 

When $B$ exceeds $B_{\mathrm{c}}$, at $\varphi=0\mathrm{mod}\,\pi$ both  $\chi_{22}$ and $\chi_{33}$ are dominated by a single peak at $q_0$ [see panel (c) from Fig.~\ref{fig: susceptiblity in field}; $q_0$ is defined below Eq.~\eqref{eq: peak in chi 11}]. This, together with the absence of such a peak in $\chi_{11}$, suggests that helical magnetic order in the 2-3 plane will form. 
At the same value of the magnetic field, but for $\varphi=\pi/2\mathrm{mod}\pi$,
a double-peak structure is eminent at the same wave vectors in $\chi_{11}$, $\chi_{33}$, and $\chi_{13}$, while $\chi_{22}$ has a double-peak at different wave vectors, while the off-diagonal elements $\chi_{12}$ and $\chi_{23}$ nearly vanish. This suggests that a helical order in the 1-3 plane will persist at $\varphi=\pi/2$. 

We can summarize these results as follows: A homogeneous magnetic field $\bs{B}$ will, in addition to inducing a homogeneous magnetization parallel to its direction, also modify the spatially varying (helical) component of the magnetic order. The critical scale that it has to exceed for qualitatively changing the helical order is given by $B_{\mathrm{c}}$ from Eq.~\eqref{eq: critical field}. If $B<B_{\mathrm{c}}$, the helical component of the magnetization remains in the 1-3 plane, pinned by Rashba spin-orbit coupling and independent of the direction of $\bs{B}$. In contrast, if $B>B_{\mathrm{c}}$, the helical component of the magnetization will lie in the plane perpendicular to $\bs{B}$, i.e., in the 2-3 plane if $\bs{B}$ points in the 1-direction and in the 1-3 plane if $\bs{B}$ points in the 2-direction.
This is summarized in the schematic phase diagram Fig.~\ref{fig: magnetic phase diagram}. For what follows, the important assumption that we derive from this phase diagram is that the helical magnetic order remains stable in the 1-3 plane for inplane magnetic fields $B<B_{\mathrm{c}}$ irrespective of their orientation $\varphi$. While the discussion above is largely speculative, the only thing we need for our setup is that the applied magnetic field influences the helix to a much less extent than it influences the induced $p$-wave gap on the chain. This is likely to be true, as the energy scale for the induced $p$-wave gap on the chain is much smaller than that of the magnetic atoms.~\cite{Nadj-Perge13}

\section{Superconducting states in presence of helical magnetic order}
\label{sec: superconductivity}

\subsection{Superconducting phase of a straight chain}

We are now going to investigate the one-dimensional electronic model of the chain subject to proximity-induced superconducting pairing, the helical magnetic order and Rashba spin-orbit coupling as well as an in-plane Zeeman field. We assume that the classical helical magnetic order discussed in the last section is rigidly formed and couples via a Hunds-type coupling to the conduction electrons. Given the helical magnetic order, the Rashba spin-orbit coupling is not a physically necessary ingredient for the effects that we are interested in, but we still include it for a more complete tight-binding Hamiltonian.
Consider the following model Hamiltonian for a straight chain of $L$ atoms
\begin{equation}
\begin{split}
H=&\,\sum_{n=1}^L
\left\{c^\dagger_n (t+\mathrm{i}\alpha\,\sigma_2)c^{\ }_{n+1}
+\Delta c^\dagger_{n,\uparrow}c^\dagger_{n,\downarrow}
+\mathrm{h.c.}
\right\}
\\
&\,+
\sum_{n=1}^L
c^\dagger_n \left[(\bs{B}+\bs{M}_n)\cdot\bs{\sigma}-\mu\right]\,c^{\ }_{n},
\end{split}
\label{eq: straight Wire Ham}
\end{equation}
 where $c^\dagger_n=(c^\dagger_{n,\uparrow},c^\dagger_{n,\downarrow})$ and $c^\dagger_{n,s}$ creates an electron of spin $s=\uparrow, \downarrow$ on site $n=1,\cdots, L$. Here, $t$ is the nearest-neighbor hopping integral, $\alpha$ parametrizes the Rashba spin-orbit coupling, $\Delta$ is the superconducting gap, $\mu$ is the chemical potential, and $\bs{B}$ is the external magnetic field. As discussed in the last section, the magnetic moment of the helical order lies in the 1-3 plane and has the spatial dependence
 \begin{equation}
 \bs{M}_n=M[\cos\,(n\, \theta+\theta_0),0,\pm\sin\,(n\, \theta+\theta_0)]^{\mathsf{T}},
 \end{equation} 
where $M$ is the overall amplitude, $\theta$ is the tilt between adjacent moments, $\pm$ stands for the two possible helicities and $\theta_0$ is a phase shift.
If the pitch $\theta$ between adjacent moments is large (such as $\pi/3$ or $\pi/4$), the choice of phase $\theta_0$ can have profound consequences on the spectrum of Hamiltonian~\eqref{eq: straight Wire Ham} (see Fig.~\ref{fig: theta0 comparison}). For small  $\theta$, in contrast, the choice of $\theta_0$ is inconsequential as the system can be (approximately, up to corrections of order $\theta_0/2\pi$) transformed to $\theta_0=0$ by an appropriately chosen translation.

\begin{figure}
\centering
\includegraphics[width=.7\textwidth]{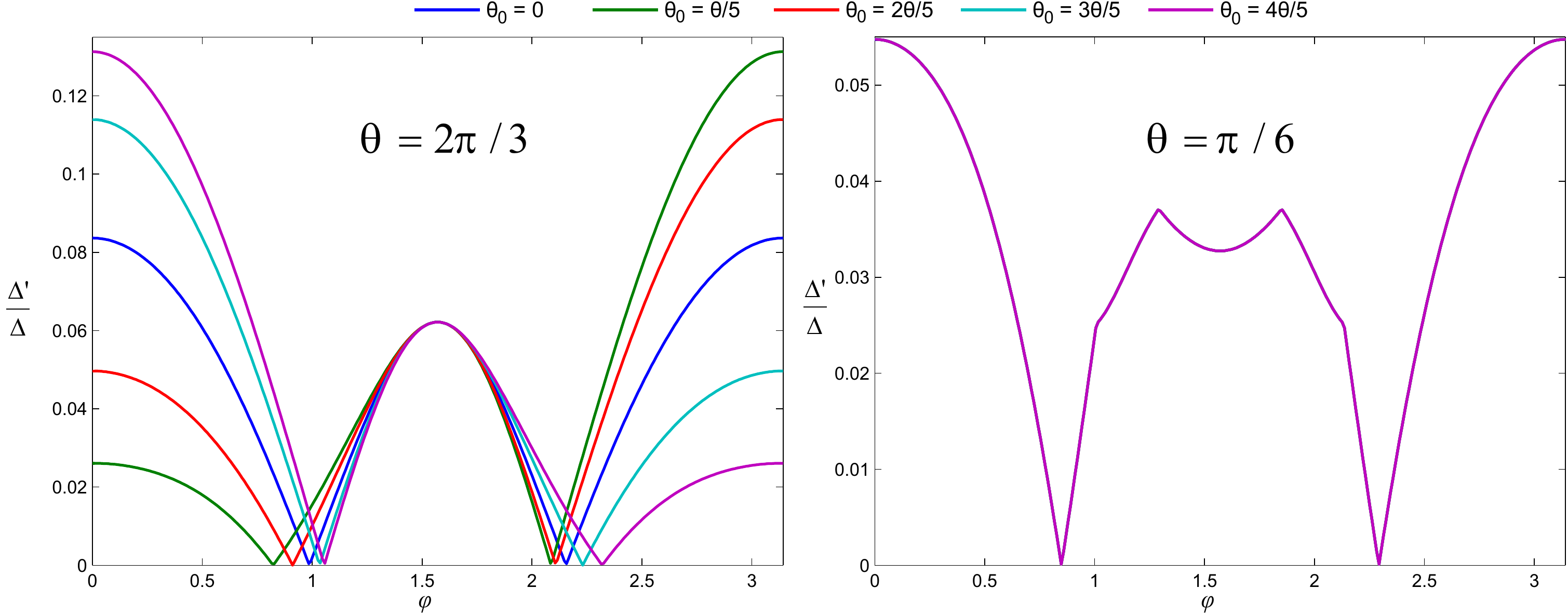}
\caption{Effects of $\theta_0$ on the bulk band gap, shown as a function of the orientation of the external $\bs{B}$-field with fixed $|\bs{B}|$ (cf. Fig.~1 of the main text). For large $\theta$ (left panel, $\theta = 2\pi/3$), the dependence of the spectra on $\theta_0$ is significant; for small $\theta$ (right panel, $\theta = \pi/6$), the dependence is almost indiscernible (all curves here fall on top of each other). The parameters used for the left panel are $\mu/\Delta=3.9$, $M/\Delta=5$, $\alpha/\Delta=0.75$, $t/\Delta=1$, $|\bs{B}|/\Delta=0.65$; the parameters used for the right panel are the same as those for Fig.~1(c) of the main text. 
}
\label{fig: theta0 comparison}
\end{figure}

Before we proceed, we shall briefly illustrate that the helical magnetic order is gauge equivalent to the Rashba spin-orbit coupling plus a homogeneous Zeeman field in 3-direction. To see that, we perform a $n$-dependent unitary rotation $U_n$ on $c_n=U_n\tilde{c}_n$ which is defined by
\begin{equation}
U^\dagger_n(\bs{M}_n\cdot\bs{\sigma}) U^{\ }_n=M\,\sigma_3
\end{equation}
and represented as
\begin{equation}
U_n=\mathrm{exp}(-\mathrm{i}n\,\theta\,\sigma_2/2).
\label{eq: gauge trafo}
\end{equation}
If the Zeeman field $\bs{B}=(0, B_2,0)$ points in the 2-direction, the Hamiltonian reads in terms of the transformed fermion operators
\begin{equation}
\begin{split}
H=&\,\sum_{n=1}^L
\left\{\tilde{c}^\dagger_n (\tilde{t}+\mathrm{i}\tilde{\alpha}\,\sigma_2)\tilde{c}^{\ }_{n+1}
+\Delta \tilde{c}^\dagger_{n,\uparrow}\tilde{c}^\dagger_{n,\downarrow}
+\mathrm{h.c.}
\right\}
+
\sum_{n=1}^L
\tilde{c}^\dagger_n \left(B_2\,\sigma_2+M\,\sigma_3-\mu\right)\,\tilde{c}^{\ }_{n},
\end{split}
\label{eq: transformed H}
\end{equation}
with 
\begin{equation}
\tilde{t}:=t\,\cos\frac{\theta}{2}+\alpha\,\sin\frac{\theta}{2},
\qquad 
\tilde{\alpha}:=\alpha\,\cos\frac{\theta}{2}-t\,\sin\frac{\theta}{2}.
\label{eq: tilde t}
\end{equation}
All spatial dependence of the matrix elements has been gauged away and traded for a renormalized hopping and Rashba spin-orbit coupling. Depending on the combination of the parameters, the gap opened by the superconducting order parameter can either be topological or trivial, in the sense that each end of the chain hosts a single Majorana bound state at zero energy or not. 

\subsubsection{Topological and trivial phase of the model}

Let us examine possible phase transitions for the model in absence of a magnetic field $\bs{B}=0$ using the representation~\eqref{eq: transformed H}.
The fact that the helical magnetic order is a Fermi surface instability (say $\theta=q_\lambda$, $\lambda=\pm$) in absence of superconducting pairing produces the following implicit equation for $\theta$
\begin{equation}
\begin{split}
0=&\,2t\,\cos\frac{\theta}{2}+\lambda2\alpha\,\sin\frac{\theta}{2}-\mu.
\end{split}
\label{eq: theta-equation}
\end{equation}
Equation~\eqref{eq: theta-equation} is the eigenvalue equation $\xi_{\theta/2}=0$ of Hamiltonian~\eqref{eq: straight Wire Ham} for $\Delta=\bs{B}=\bs{M}_n=0$, where $\xi_{k}$ is the eigenvalue at momentum $k$ and the condition $\xi_k=0$ determines the Fermi momenta. It is the lattice analogue of Eq.~\eqref{eq: peak in chi 11}.
Here, we considered the unrealistic but analytically tractable situation of the 1D RKKY interaction as the origin of the helix. We note that in the realistic 2D RKKY case \cite{Kim14}, where the helix configuration depends on the Rashba spin-orbit coupling, tantamount but similar physics takes place. The dispersion of Hamiltonian~\eqref{eq: transformed H} reads for $B_2=0$
\begin{equation} \label{eq:sp0}
\begin{split}
\tilde{\xi}_k
=&\,\pm\sqrt{
M^2+\Delta^2+C_k^2+S_k^2
\pm
2\sqrt{
C_k^2S_k^2
+M^2\left(
\Delta^2+C_k^2
\right)
}
}
,
\end{split}
\end{equation}
where $C_k\equiv2\tilde{t}\cos\,k-\mu$ and $S_k\equiv2\tilde{\alpha}\sin\,k$. At $k=0$, the effective Hamiltonian takes a Dirac form for two dispersing modes with masses  
\begin{equation}
\begin{split}
\tilde{\xi}_0
=&\,\pm M\pm\sqrt{\Delta^2+(2\tilde{t}-\mu)^2
}
.
\end{split}
\end{equation}
There is a topological phase transition when either of these masses changes sign, namely at $|M|=\sqrt{\Delta^2+(2\tilde{t}-\mu)^2}$.

In contrast, at $k=\pi$, the energies are
\begin{equation}
\begin{split}
\tilde{\xi}_\pi
=&\,\pm
M
\pm
\sqrt{\Delta^2+(2\tilde{t}+\mu)^2
}
,
\end{split}
\end{equation}
and hence there is a phase transition at $|M|=\sqrt{\Delta^2+(2\tilde{t}+\mu)^2}$. Assuming without loss of generality that $|2\tilde{t}-\mu|<|2\tilde{t}+\mu|$, this yields the following condition for being in the \emph{topological phase}:
\begin{equation}
\sqrt{\Delta^2+(2\tilde{t}-\mu)^2}
<M<
\sqrt{\Delta^2+(2\tilde{t}+\mu)^2}.
\label{eq: condition for topological}
\end{equation}

\subsubsection{Topological phase transition with $B_2$}\label{sssec:tpt_B2}

As the Rashba spin-orbit coupling, the magnetic field $B_2$ enters the Hamiltonian by multiplying the second Pauli matrix. In effect, $B_2$ thus shifts $\sin\,k\to -B_2/\alpha+\sin\,k$, removing the spectral degeneracy between $k$ and $-k$. Since the formation of Cooper pairs is energetically disfavored if the participating electrons are not at momenta $k$ and $-k$, $B_2$ will generically drive a phase transition into a gapless state (with indirect band gap), iff
\begin{equation}
|B_2|>|\Delta|\qquad\text{gapless}
\label{eq: gaplessness condition}
\end{equation}
independent of $M$, as long as $|M|<|2\tilde{t}+\mu|$. Here, we assume $|2\tilde{t}-\mu|<|2\tilde{t}+\mu|$ and $(2\tilde{t}-\mu)(2\tilde{t}+\mu)>0$.

To see how Eq.~\eqref{eq: gaplessness condition} comes about, we note that the condition for a zero-energy eigenvalue of Hamiltonian~\eqref{eq: transformed H} is given by
\begin{equation}\label{eq:det_ham_B2}
D(k)=(S_k^2 + C_k^2 - M^2 + \Delta^2 - B_2^2)^2 - 4 S_k^2 (C_k^2 - M^2)=0,
\end{equation}
where $C_k\equiv2\tilde{t}\cos\,k-\mu$ and $S_k\equiv2\tilde{\alpha}\sin\,k$ are the same as in Eq.~\eqref{eq:sp0}. The existence of a solution to this equation requires
\begin{align}
  &S_k =0,\; C_k^2 - M^2 + \Delta^2 - B_2^2 = 0, \label{eq:E0solutions_case1}\\
  \text{or}\quad &S_k \ne 0,\;  C_k^2 - M^2 \ge 0,\;
  \Bigl(S_k \pm \sqrt{C_k^2 - M^2}\Bigr)^2 = B_2^2 - \Delta^2.
\end{align}

Let us first focus on the latter case, where $k\ne 0$ or $\pi$, and the solutions always come in pairs as $\pm k$. Clearly a solution in this case can exist only if $B_2^2 - \Delta^2\ge0$. When $B_2^2 - \Delta^2=0$, as long as $|M|<|2\tilde{t}+\mu|$, we find that $|S_k|-\sqrt{C_k^2 - M^2}< 0$ when $k$ approaches $\pi$, and  $|S_k|-\sqrt{C_k^2 - M^2}> 0$ when $k$ approaches the point of $k$ where $C_k^2 - M^2=0$, which implies that at least one pair of solutions of the original equation exist. By the same token, when $(2\tilde{t}+\mu)^2-M^2>B_2^2 - \Delta^2>0$, solutions of $k\ne 0$ or $\pi$ still exist to the original equation \eqref{eq:det_ham_B2}. This proves the statement of Eq.~\eqref{eq: gaplessness condition}.

In the case of Eq.~\eqref{eq:E0solutions_case1}, $k = 0$ or $\pi$, and the condition $|B_2|>|\Delta|$ is not necessary. Instead the existence of a zero-energy solution requires $\tilde{\xi}_0=0$ or $\tilde{\xi}_\pi=0$ with
\begin{equation}
\begin{split}
&\tilde{\xi}_0=\pm\left[\sqrt{B_2^2+M^2}-\sqrt{\Delta^2+(2\tilde{t}-\mu)^2}\right], \\
&\tilde{\xi}_\pi=\pm\left[\sqrt{B_2^2+M^2}-\sqrt{\Delta^2+(2\tilde{t}+\mu)^2}\right].
\end{split}
\end{equation}
That is, a direct gap closing is induced by the field $B_2$ at $k=0$ or $\pi$, if 
\begin{equation}
B_2^2 + M^2 = \Delta^2 + (2\tilde{t}\mp\mu)^2.
\label{eq: special values M}
\end{equation}

The regime that we are interested in is $|M|>|\Delta|,|B_2|$. In view of Eq.~\eqref{eq: condition for topological}, no gap closing phase transition out of the topological phase can be induced that corresponds to $\tilde{\xi}_0=0$ by ramping up $B_2$. However, for suitable choice of chemical potential $\mu$, $B_2$ can induce a topological phase transition at $\tilde{\xi}_\pi=0$, while maintaining the condition $|B_2|<|\Delta|$ to escape entering the gapless phase. To sum up, for the parameters relevant to our proposal, we start from a topological phase by satisfying condition~\eqref{eq: condition for topological} and ramp up $B_2$ until one of the two things happen:
Either, if $|M|>|2\tilde{t}+\mu|$, we enter a gapped topologically trivial phase at $B_2=\sqrt{\Delta^2+(2\tilde{t}+\mu)^2-M^2}$.
Or, if $|M|<|2\tilde{t}+\mu|$, we enter a gapless phase as soon as $|B_2|\geq|\Delta|$.

\subsubsection{Topological phase transition with $B_1$}

To study the effect of a homogeneous field $B_1$ in the 1-direction, the gauge transformation~\eqref{eq: gauge trafo} is only partially useful, for it does not generate a Hamiltonian with the translational symmetry of the lattice.
However, we can use it to move the position dependence from a large term in the Hamiltonian (i.e., $M$) to a small one (i.e., $B_1$). The transformed Hamiltonian reads
\begin{equation}
\begin{split}
\tilde{H}=&\,\sum_{n=1}^L
\left\{\tilde{c}^\dagger_n (\tilde{t}+\mathrm{i}\tilde{\alpha}\,\sigma_2)\tilde{c}^{\ }_{n+1}
+\Delta \tilde{c}^\dagger_{n,\uparrow}\tilde{c}^\dagger_{n,\downarrow}
+\mathrm{h.c.}
\right\}
+
\sum_{n=1}^L
\tilde{c}^\dagger_n \left(\tilde{\bs{B}}_n\cdot\bs{\sigma}+M\,\sigma_3-\mu\right)\,\tilde{c}^{\ }_{n},
\end{split}
\end{equation}
where the definitions of $\tilde{t}$ and $\tilde{\alpha}$ carry over from Eq.~\eqref{eq: tilde t} and 
 \begin{equation}
 \tilde{\bs{B}}_n=B_1[\cos\,(n\, \theta+\theta_0),0,\pm\sin\,(n\, \theta+\theta_0)]^{\mathsf{T}},
 \end{equation} 
with $\theta_0=\pi/2$.

We can study it in the limit where both $B_1$ and $\Delta$ are small compared to the other energy scales. We will find that $B_1$ and $\Delta$ parametrize competing mass gaps of Hamiltonian~\eqref{eq: transformed H}. If $\Delta$ ($B_1$) dominates, the system is in the topological (trivial) phase. 
In absence of $B_1$, the Bloch Hamiltonian that corresponds to Eq.~\eqref{eq: transformed H} in the basis $(c_{k,\uparrow},c_{k,\downarrow},c^\dagger_{-k,\uparrow},c^\dagger_{-k,\downarrow})$ is given by
\begin{equation}
\mathcal{H}(k)
=
\begin{pmatrix}
2\tilde{t} \cos\,k-\mu+M&-2\mathrm{i}\tilde{\alpha}\sin\,k&0&\Delta\\
2\mathrm{i}\tilde{\alpha}\sin\,k&2\tilde{t}\cos\,k-\mu-M&-\Delta&0\\
0&-\Delta&-2\tilde{t}\cos\,k+\mu-M&2\mathrm{i}\tilde{\alpha}\sin\,k\\
\Delta&0&-2\mathrm{i}\tilde{\alpha}\sin\,k&-2\tilde{t}\cos\,k+\mu+M
\end{pmatrix}.
\label{eq: Bloch Hamiltonian}
\end{equation}
Linearized around the Fermi momentum $k_{\mathrm{F}}$ in absence of the superconducting gap (but with nonzero $M$), the Bloch Hamiltonian takes the general form
\begin{equation}
\mathcal{H}_{\mathrm{lin},\Delta}(p)
=
\begin{pmatrix}
F+\frac{\tilde{t}}{\tilde{\alpha}}\sqrt{F G}\,p&\mathrm{i}\sqrt{F G}+\mathrm{i}Lp&0&\Delta\\
-\mathrm{i}\sqrt{F G}-\mathrm{i}Lp&G+\frac{\tilde{t}}{\tilde{\alpha}}\sqrt{F G}\,p&-\Delta&0\\
0&-\Delta&-F-\frac{\tilde{t}}{\tilde{\alpha}}\sqrt{F G}\,p&-\mathrm{i}\sqrt{F G}-\mathrm{i}Lp\\
\Delta&0&\mathrm{i}\sqrt{F G}+\mathrm{i}Lp&-G-\frac{\tilde{t}}{\tilde{\alpha}}\sqrt{F G}\,p
\end{pmatrix}.
\end{equation}
Here, $k=k_{\mathrm{F}}+p$ and $F$, $G$, and $L$ are constants that depend on the parameters in Eq.~\eqref{eq: Bloch Hamiltonian}.
To linear order in $\Delta$, the gap $\Delta_{\mathrm{lin}}$ of $\mathcal{H}_{\mathrm{lin}}(p)$ at $p=0$ is given by
\begin{equation}
\begin{split}
\Delta_{\mathrm{lin}}
=&\,
\Delta\sqrt{\left|1-\frac{(F-G)^2}{(F+G)^2}\right|}
\\
=&\,
\Delta\sqrt{\left|1-\frac{4M^2(\tilde{t}^2+\tilde{\alpha}^2)^2}{\left[\tilde{\alpha}^2\mu+\tilde{t}\,
\sqrt{(M^2+4\tilde{\alpha}^2)(\tilde{t}^2+\tilde{\alpha}^2)-\tilde{\alpha}^2\mu^2}\right]^2}\right|}\ .
\end{split}
\end{equation}

This has to be contrasted with the gap that is introduced by $B_1$. In this case, the effective Hamiltonian reads
in the basis $(c_{k_{\mathrm{F}}+p,\uparrow},c_{k_{\mathrm{F}}+p,\downarrow},c_{k_{\mathrm{F}}+\theta+p,\uparrow},c_{k_{\mathrm{F}}+\theta+p,\downarrow})$ 
\begin{equation}
\mathcal{H}_{\mathrm{lin},B_1}(p)
=
\begin{pmatrix}
F+\frac{\tilde{t}}{\tilde{\alpha}}\sqrt{F G}\,p&\mathrm{i}\sqrt{F G}+\mathrm{i}Lp&\mathrm{i}B_1/2&B_1/2\\
-\mathrm{i}\sqrt{F G}-\mathrm{i}Lp&G+\frac{\tilde{t}}{\tilde{\alpha}}\sqrt{F G}\,p&B_1/2&-\mathrm{i}B_1/2\\
-\mathrm{i}B_1/2&B_1/2&F-\frac{\tilde{t}}{\tilde{\alpha}}\sqrt{F G}\,p&\mathrm{i}\sqrt{F G}-\mathrm{i}Lp\\
B_1/2&\mathrm{i}B_1/2&-\mathrm{i}\sqrt{F G}+\mathrm{i}Lp&G-\frac{\tilde{t}}{\tilde{\alpha}}\sqrt{F G}\,p
\end{pmatrix}.
\end{equation}
To linear order in $B_1$, the gap that emerges at $p=0$ is given by
\begin{equation}
\begin{split}
\Delta_{B_1}
=&\frac{B_1}{2} \left|\frac{F-G}{F+G}\right|
\\
=&\frac{B_1}{2} \left|
\frac{M(\tilde{t}^2+\tilde{\alpha}^2)}{\tilde{\alpha}^2\mu+\tilde{t}\,
\sqrt{(M^2+4\tilde{\alpha}^2)(\tilde{t}^2+\tilde{\alpha}^2)-\tilde{\alpha}^2\mu^2}}\right|\ .
\end{split}
\end{equation}
As $\Delta_{B_1}$ and $\Delta_{\mathrm{lin}}$ are acting on different degrees of freedom (electrons with momenta differing by $\theta$ in one case and the electron-hole space in the other case), they are commuting and hence competing mass terms in an effective Dirac equation. Hence, the larger of $\Delta_{B_1}$ and $\Delta_{\mathrm{lin}}$ determines the phase of the system. If $\Delta_{\mathrm{lin}}>\Delta_{B_1}$, the superconductor is topological, hosting Majorana end states. If $\Delta_{\mathrm{lin}}<\Delta_{B_1}$ it is trivial.
Importantly, if parameters are such that
\begin{equation}
\left|\frac{F-G}{F+G}\right|\sim1,
\end{equation}
the gap $\Delta_{B_1}$ scales faster with $B_1$ than the gap $\Delta_{\mathrm{lin}}$ scales with $\Delta$. 
As a consequence, for some given superconducting gap $\Delta$, we can expect a phase transition from the topological to the trivial superconductor at a critical field $B_{1,\mathrm{c}}$ which is 
\begin{equation}
B_{1,\mathrm{c}}<\Delta.
\end{equation}

\subsubsection{Phase diagram}
We have now sown that both $B_1$ and $B_2$ can induce a topological phase transition in the chain, based on very different mechanisms. 
Crucially, the critical field strength $B_{1,\mathrm{c}}$ at which this phase transition appears is different from the critical field strength $B_{2,\mathrm{c}}$.
For appropriate choice of the parameters,
the system enters a gapless phase at $B_2=B_{2,\mathrm{c}}=\Delta$, while a transition from the topological to the gapless phase triggered by $B_1$ can happen at $B_1=B_{1,\mathrm{c}}<\Delta$.
This has the following consequence: If the field strength of the external field $B$ is chosen such that $B_{1,\mathrm{c}}<B<B_{2,\mathrm{c}}$, then by rotating the external field in the 1-2 plane
\begin{equation}
\bs{B}=B(\cos\varphi,\sin\varphi,0)^{\mathsf{T}},
\label{eq: Bfield parametrization}
\end{equation}
the chain has to undergo a topological-to-trivial phase transition for some $\varphi=\varphi_0$ between $0$ and $\pi/2$ (see Fig.~2 from the main text). 
At this transition the bulk of the chain becomes gapless and two Majorana states emerge. This phase transition occurs while the field remains in the 1-2 plane. It is therefore not harmful to the superconducting substrate if this is a thin film. 

\subsection{Topological transitions in the Rashba wire}

Similar to the system with magnetic helix, also a straight wire with a suitable combination of Rashba spin-orbit coupling and magnetic field can feature a topological phase transition as a function of the orientation of the magnetic field. Let us consider the following Hamiltonian
\begin{equation}
\begin{split}
\tilde{H}=&\,\sum_{n=1}^L
\left\{c^\dagger_n (t+\mathrm{i}\alpha\,\sigma_2)c^{\ }_{n+1}
+\Delta c^\dagger_{n,\uparrow}c^\dagger_{n,\downarrow}
+\mathrm{h.c.}
\right\}
+
\sum_{n=1}^L
c^\dagger_n \left(\bs{B}\cdot\bs{\sigma}-\mu\right)\,c^{\ }_{n}.
\end{split}
\label{eq:  H for Rashba wire}
\end{equation}
Formally this Hamiltonian is different from Hamiltonian~\eqref{eq: transformed H} only in that the magnetic ingredient is constant in space for the former, but position-dependent for the latter.

\begin{figure}
\centering
\includegraphics[width=.8\textwidth]{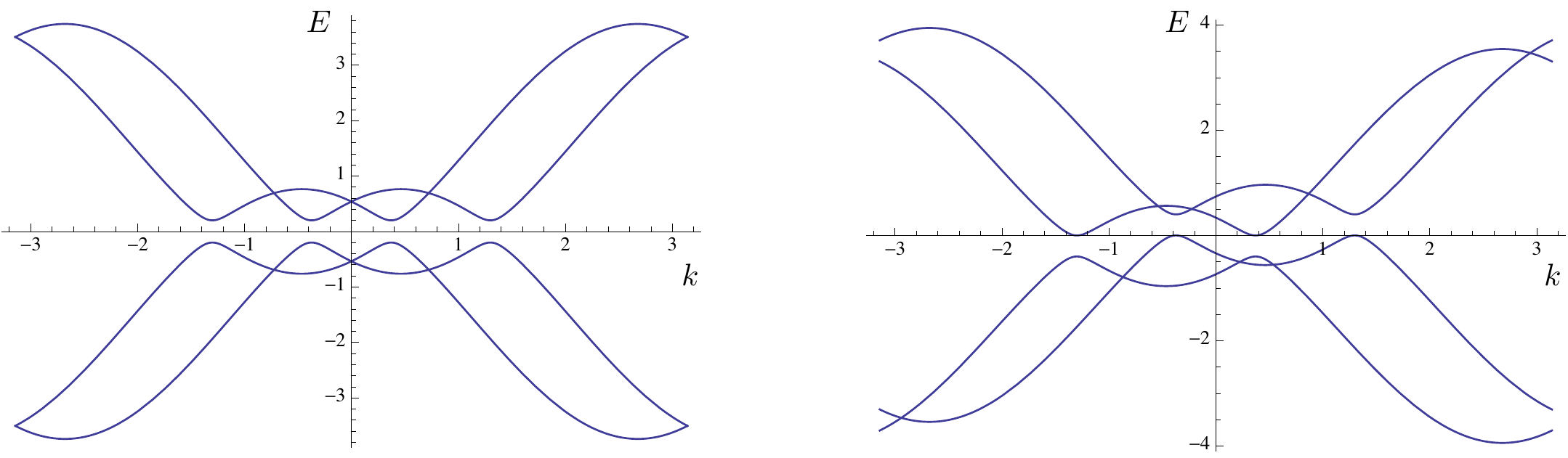}
\caption{Spectrum of Hamiltonian~\eqref{eq:  H for Rashba wire} for $\mu/t=1.5$, $\Delta/t=0.2$, $\mu/t=0.5$, as well as $\bs{B}=0$ (left panel) and $\bs{B}=(0,0.2,0)$ (right panel). For the parameter values on the right panel, i.e., $|B_2|=|\Delta|$, the (indirect) superconducting gap closes.
}
\label{fig: rashba wire spectrum}
\end{figure}

Hamiltonian~\eqref{eq:  H for Rashba wire} is gapless for $|B_2|>|\Delta|$, which follows the same reason as we have shown in Sec.~\ref{sssec:tpt_B2} for Eq.~\eqref{eq: gaplessness condition}. The gap that closes at $|B_2|=|\Delta|$ is an indirect gap for generic parameters (see Fig.~\ref{fig: rashba wire spectrum}).
If gapped, the ground state of Hamiltonian~\eqref{eq:  H for Rashba wire}
is topological if
\begin{equation}
 \min(A_-,A_+)<|\bs{B}|<  \max(A_-,A_+),
\end{equation}
where $A_{\pm}:=\sqrt{(2t\pm\mu)^2+\Delta^2}>|\Delta|$.

We want to study topological phase transitions of Hamiltonian~\eqref{eq:  H for Rashba wire} as a function of $\bs{B}$ between gapped phases. For that reason, we will always require $|B_2|<|\Delta|$. As long as this condition is met, owing to particle-hole symmetry, a topological phase transition can only happen when a direct gap closes and reopens at the momenta $k=0$ or $\pi$. These momenta are also inversion symmetric momenta at which the spin-orbit coupling term is vanishing. Therefore the Bloch Hamiltonian in the Nambu basis has the particularly simple form
\begin{equation}
\tilde{\mathcal{H}}(k=0,\pi)
=
[(2t\cos{k}-\mu)\tau_3+\Delta\tau_1]\otimes\sigma_0 + \tau_0\otimes(\bs{B}\cdot\bs{\sigma}),
\label{eq: Bloch Hamiltonian for Rashba wire}
\end{equation}
where $\tau$'s stand for the Pauli matrices for the particle-hole components. The spectrum of this Hamiltonian at $k=0,\pi$ only depends on $|\bs{B}|$ instead of the orientation of $\bs{B}$, because for $\bs{B}=0$ the Hamiltonian~\eqref{eq: Bloch Hamiltonian for Rashba wire} has the full $SU(2)$ spin-rotation symmetry.

\begin{figure}
\centering
\includegraphics[width=.6\textwidth]{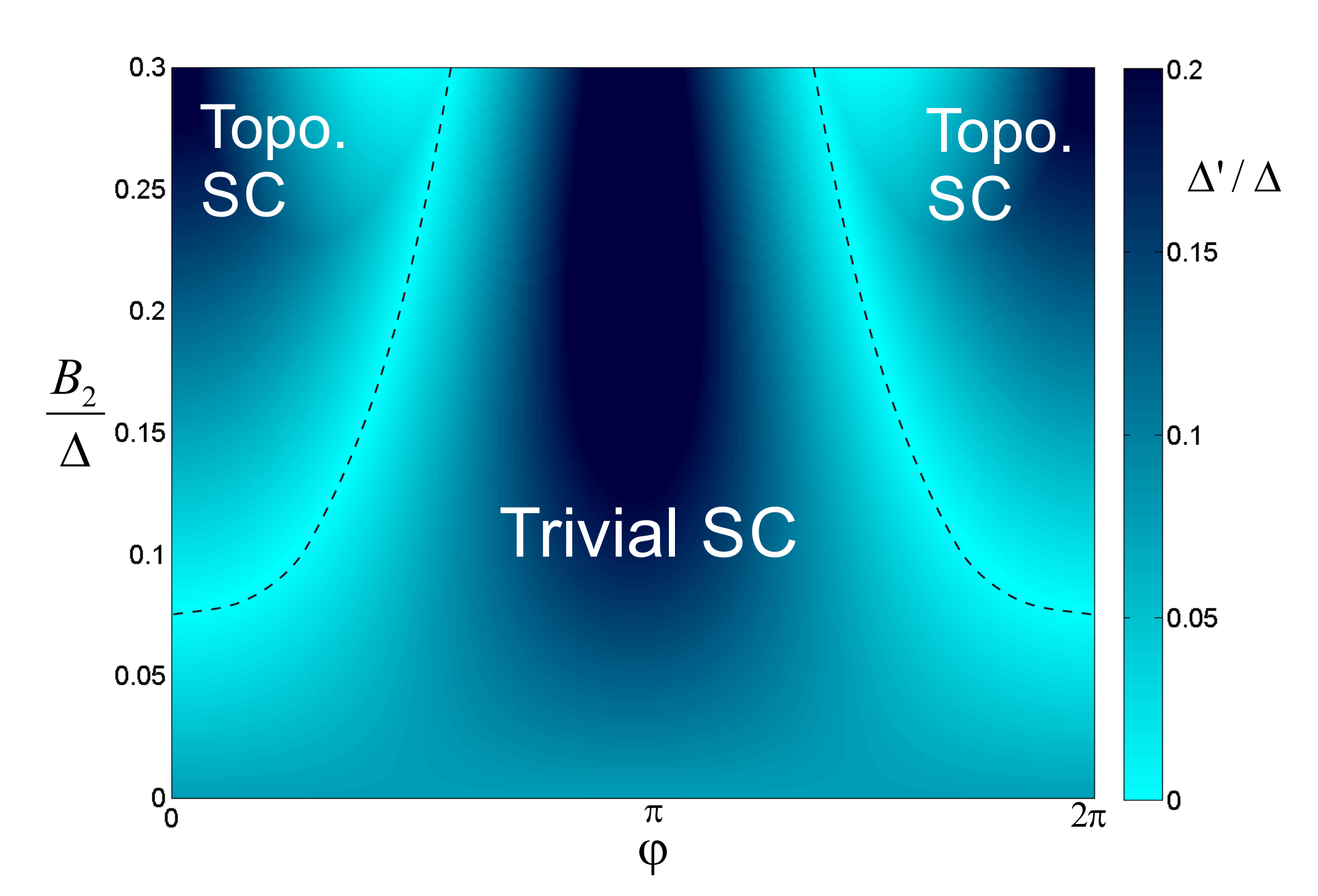}
\caption{Phase diagram of a Rashba wire in a combined magnetic field $\bs{B} = (B_1+B_2\cos\varphi, B_2\sin\varphi, 0)$ with $B_1 = 0.24$. The other parameters are $t=-1$, $\mu=-1.9$, $\alpha=0.4$ and $\Delta=0.3$. Note in particular that here $\varphi$ varies from $0$ to $2\pi$ whereas in Fig.~1(b) of the main text $\varphi$ varies from $0$ to $\pi$.
}
\label{fig: rashba wire phase diagram}
\end{figure}

The eigenvalues of \eqref{eq: Bloch Hamiltonian for Rashba wire} are given by
\begin{equation}
\xi(k=0)=\pm |\bs{B}| \pm A_-,
\qquad
\xi(k=\pi)=\pm |\bs{B}| \pm A_+,
\end{equation}
where the signs are uncorrelated. Assuming a weak magnetic field, the direct gap closes when $|\bs{B}| = \min(A_-,A_+)$, irrespective of the orientation of the field. This immediately rules out the possibility of a topological phase transition induced by varying the orientation of a magnetic field while keeping its magnitude constant. However, this can still be achieved by applying a constant background magnetic field on top of the rotating one. For example, if $\bs{B} = (B_1+B_2\cos\varphi, B_2\sin\varphi, 0)$, with $|B_1|+|B_2|>\min(A_-,A_+)>\sqrt{B_1^2+B_2^2}$, two topological phase transitions will occur when $\varphi$ is tuned from $0$ to $2\pi$ (see Fig.~\ref{fig: rashba wire phase diagram}).

\subsection{Spectral properties of a circle and finite size scaling}

We now return to consider the chain with helical magnetic order.
In order to move Majorana bound states in a controlled way, one would like to pin them to some mobile domain wall that marks the phase transition between a trivial and a topological state. The external field is fixed in angle and magnitude. However, one can utilize the directional dependence and obtain a domain wall between two chain segments that are at an angle different from $\pi$ to one another. Depending on that angle and the field orientation, one segment could be in the topological, the other in the trivial phase. In this case the two host a Majorana state between them.  More generally, one can join up many such segments into a bent chain or a circle. We assume the rotation plane of the helix is locally unaffected by the bending. That is, the magnetic moments locally lie in the plane that is spanned by the normal of the superconductor and the tangent to the bent chain.  
For the latter, the angle $\varphi$ that enters the Hamiltonian~\eqref{eq: straight Wire Ham} via the parametrization~\eqref{eq: Bfield parametrization} becomes position dependent as $\varphi\to\varphi_n=2\pi n/L $ and the boundary conditions are changed from open to periodic. To retain the same helical magnetic order in the circle as in the straight chain, the limit in which the radius of the circle is much larger than the wave length of the magnetic helix, i.e., $2\pi/L\ll\theta$, has to be assumed.

\begin{figure}[t]
\centering
\includegraphics[width=.74\textwidth]{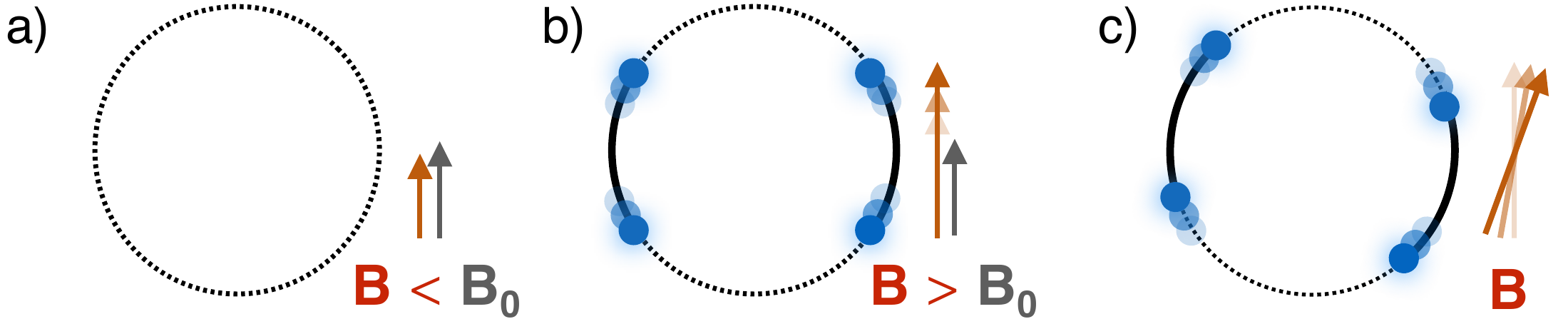}
\caption{
Majorana zero energy bound states on a circle.
(a) If the external magnetic field $B$ is below a threshold value $B_{1,\mathrm{c}}$ the entire circle is in a topological superconducting state. 
(b) Once $B$ is in between $B_{1,\mathrm{c}}$ and $B_{2,\mathrm{c}}$, two trivial superconducting segments open up at opposite end of the circle. Each interface between a trivial and a topological segment hosts a Majorana bound state (blue dot). 
(c) The position of the bound states depends on the field orientation. The Majorana zero energy modes can thus be moved by rotating the external magnetic field. Notice that each of the 4 Majorana states braids once around each other Majorana state during a $2\pi$ rotation of the external magnetic field. 
}
\label{fig: Majorana circle}
\end{figure}

As shown schematically in Fig.~\ref{fig: Majorana circle}, the circle is then capable of hosting 4 Majorana bound states. These bound states can be moved by changing the magnetic field orientation. One potential issue of this setup regards other possible bound states of the domain walls at nonzero energy, besides the topological Majorana bound states. Such excited domain wall states can be close in energy to the Majorana bound states in a large circle, and will be the major threat to an adiabatic operation on the Majorana qubits. Therefore it is crucial to compare the scaling behavior of the Majorana splitting energy ($E_M$) and that of the first (domain-wall) excitation energy ($E_\text{ex.}^{(1)}$). Straightforward numerical diagonalization of the problem shows that $E_M$ decays exponentially with an increasing circumference $L$ [see Fig.~\ref{fig:scaling}(a)], whereas $E_\text{ex.}^{(1)}$ changes in a much slower rate as $1/\sqrt{L}$ [see Fig.~\ref{fig:scaling}(b)]. This drastic difference is due to the different origin of these two energies: $E_M$ is a result of a finite overlap between the exponential tails of different Majorana wave-functions, while $E_\text{ex.}^{(1)}$ is related to the \textit{local} spatial profile of the domain wall. In the next subsection we will analyze these behaviors in more detail by using an effective model for the domain wall. It is important to notice that in a circle of several hundreds of sites, $E_M\sim 10^{-5}\Delta$ while $E_\text{ex.}^{(1)}>10^{-2}\Delta$, which leaves enough room for time/energy scales of quasi-adiabatic processes in between.

\begin{figure}[t]
\centering
\includegraphics[width=.96\textwidth]{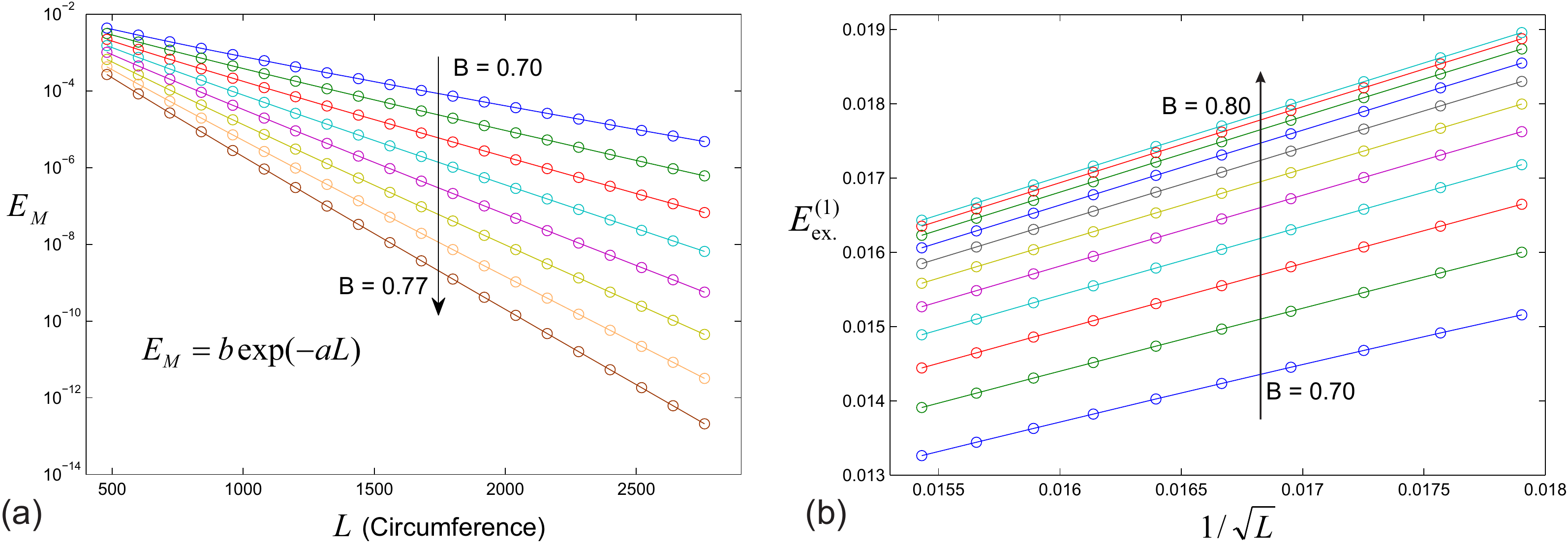}
\caption{
Finite size scaling of the low-energy domain wall states in the circle geometry.
(a) The energy-splitting of the Majorana bound states scales exponentially in $L$.
(b) The energy of the first excited state above the Majorana bound states scales as $1/\sqrt{L}$. Cases are compared with varying external magnetic field $B$ and otherwise the same parameters as in Fig. 2 of the main text.
}
\label{fig:scaling}
\end{figure}


\subsection{Domain wall states in a circle}

\begin{figure}
\centering
\includegraphics[width=.4\textwidth]{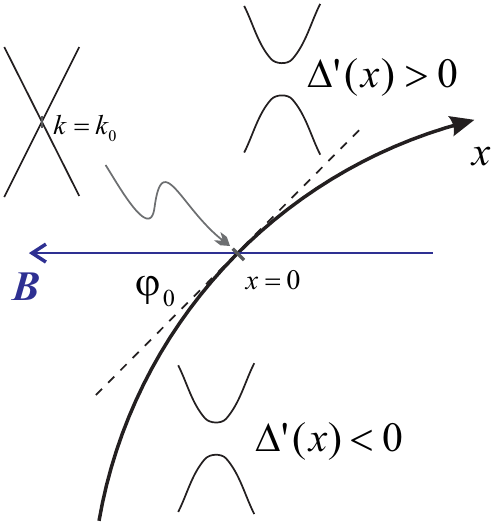}
\caption{
Illustration of the effective model Eq.~\eqref{eq:ham0} describing a domain wall on a circle.
}
\label{fig: Domain Wall Model}
\end{figure}

In this section, we would like to study the low-energy excitations above the Majorana states in an effective model that can be treated analytically. We anticipate that these lowest excitations at finite energy are localized at the same position as the Majorana bound states. It is convenient to introduce a continuous coordinate 
\begin{equation}
x:=\frac{\varphi-\varphi_0}{2\pi}L,
\end{equation}
where $L$ is the circumference of the circle and $\varphi$ is angular coordinate measured relative to the orientation of the external $\bs{B}$ field. As discussed before, the Majorana state is localized at the critical angle $\varphi_0$, which corresponds to $x=0$. We work in the continuum limit in which we can formulate an effective theory in the coordinate $x$ instead of the lattice site labels $n$, provided that $L$ is much larger than all other length scales (in particular the wavelength of the helical magnetic order) of the chain. In this limit, the generic low-energy effective Hamiltonian for excitations near $x=0$ reads (see Fig.~\ref{fig: Domain Wall Model})
\begin{align}\label{eq:ham0}
  H =
  \begin{pmatrix}
    -\hbar v (i\partial_x + k_0) & \Delta'(x) \\
    \Delta'(x) & \hbar v (i\partial_x + k_0)
  \end{pmatrix}.
\end{align}
We assume around the domain wall, the low-energy physics is dominated by states of momenta close to $k_0$, namely the momentum where the bulk gap closes and reopens at field angle $\varphi_0$ in the straight chain. We linearize the dispersion relation at the domain wall center, with respect to $k=k_0$, with a group velocity $v$. Finally, in writing Hamiltonian~\eqref{eq:ham0}, we assumed that the gap parameter $\Delta'$ around the domain wall changes linearly with respect to the angle $\varphi$ between the tangent line of the circle and the magnetic filed direction [cf. Fig.~\ref{fig:EMdeclen}(a)]. Translated to the local coordinate we have chosen, this means $\Delta'(x) = sx/r$, where $s=\partial{\Delta'}/\partial{\varphi}$ and $r=L/2\pi$ is the radius of the circle.

Let us set $\hbar=1$, and choose $k_0=0$ for the moment. The above Hamiltonian can be rewritten and transformed into a more convenient form
\begin{align}\label{eq:ham1}
  H = \sqrt{sv/r}
  \begin{pmatrix}
    & \tilde{x}-\partial_{\tilde{x}} \\
    \tilde{x}+\partial_{\tilde{x}} &
  \end{pmatrix}, \quad
  \tilde{x} \equiv \sqrt{\frac{s}{vr}}x.
\end{align}
The solutions to this Hamiltonian are well-known in the context of graphene. The same Hamiltonian applies to the Landau level problem of a single species of Dirac electrons subject to an out-of-plane magnetic field. By this analogy, $s/r$ is equivalent to the strength of the magnetic field.

The Hamiltonian \eqref{eq:ham1} has one zero energy eigenstate, the Majorana bound state, given by
\begin{align}
  \Psi_0(\tilde{x}) \propto \exp(-\tilde{x}^2/2)
  \begin{pmatrix}
    1 \\ 0
  \end{pmatrix}.
\end{align}
In the original coordinate, this solution becomes
\begin{align}
  \Psi_0(x) =
  \left(\frac{s}{\pi vr}\right)^{1/4}\exp(-\frac{s}{2vr}x^2)
  \begin{pmatrix}
    1 \\ 0
  \end{pmatrix},
\end{align}
where the wavefunction has been normalized in $(-\infty,+\infty)$.

\begin{figure}
  \centering
  \includegraphics[width=0.96\textwidth]{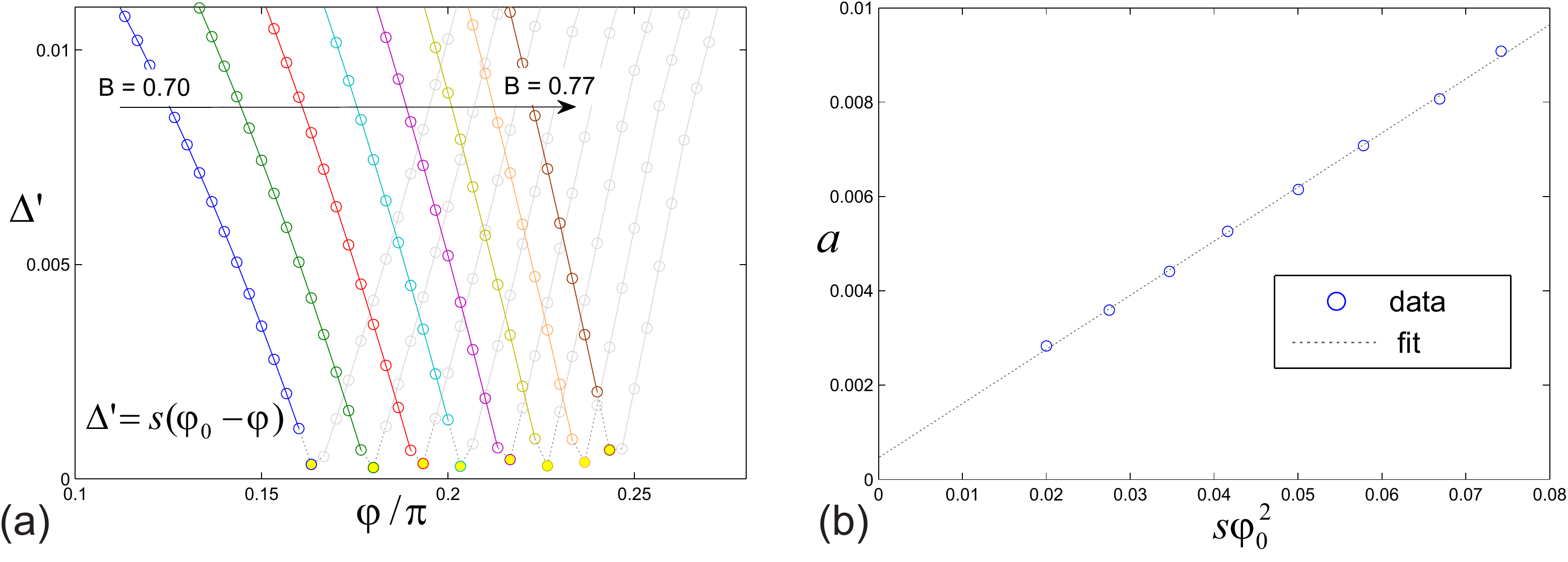}
  \caption{(a) Bulk band gaps $\Delta'$ in the vicinity of topological phase transitions. (b) Test of the correlation predicted by Eq.~\eqref{eq:EM}.}
  \label{fig:EMdeclen}
\end{figure}

The Majorana coupling energy $E_M$ is then given proportional to
\begin{align}\label{eq:EM}
  E_M \propto \sqrt{\frac{s}{\pi vr}}\int_{-\infty}^{+\infty} e^{-\frac{s}{2vr}[x^2+(x-\varphi_0 r)^2]} dx = e^{-\frac{s}{4v}\varphi_0^2 r},
\end{align}
where $\varphi_0$ is the critical angle for the topological phase transition. Here, we assume without loss of generality $\varphi_0<\pi/4$. Equation~\eqref{eq:EM} shows that $E_M$ indeed decays exponentially with an increasing size of the circle. Moreover, it provides a prediction for the decay length, which is examined by numerics as follows. First we extract $s$ and $\varphi_0$ by using a linear fit to bulk band spectra as shown in Fig.~\ref{fig:EMdeclen}(a), where we have fixed all parameters but the magnitude of the external magnetic field $B$. Next we extract the corresponding inverse decay length $a$ from Fig.~\ref{fig:scaling}(a). Then we plot the correlation between $a$ and $s\varphi_0^2$ in Fig.~\ref{fig:EMdeclen}(b). Clearly, there is a remarkable agreement between the numerical results and the prediction of the effective model.

Another important length scale that can be derived from the effective model is the oscillation period of the Majorana wave-functions. This simply follows from the fact that the solutions of the Schr\"{o}dinger equation when $k_0\ne 0$ can be obtained from those when $k_0=0$ by substitution $\Psi\rightarrow\Psi\exp(ik_0 x)$. In the system we are considering, particle-hole symmetry dictates $k_0$ to be $0$ or the Brillouin zone boundary $\pm \theta/2$, because otherwise $k_0$ must appear in pairs and the system will not undergo a topological phase transition; the specific mechanism for the phase transition that we have discussed further limits $k_0$ down to $\pm \theta/2$. In this case the solutions with factors $\exp(\pm i\theta x/2)$ must be superimposed to yield $\sin(\theta x/2)$ or $\cos(\theta x/2)$. Therefore the oscillation period for the wave function amplitude is $4\pi/\theta$, while the period of the probability is $2\pi/\theta$.

In fact, all the excited-state solutions to Hamiltonian \eqref{eq:ham1} can be readily obtained. Of particular importance to us, concerning the adiabaticity of the manipulations on the Majorana states, is the first excited state
\begin{align}
  E_\text{ex.}^{(1)} = \sqrt{2sv/r}, \quad
  \Psi_\text{ex.}^{(1)}(\tilde{x}) \propto \exp(-\tilde{x}^2/2)
  \begin{pmatrix}
    \sqrt{2}\tilde{x} \\ 1
  \end{pmatrix}.
\end{align}
Evidently, its energy scales as $1/\sqrt{r}$, which is analogous to the $\sqrt{B}$ dependence of the Landau levels in graphene. This energy scaling behavior is a local property of the domain wall, in contrast to the nonlocal nature of the Majorana coupling energy, and will dominate as long as the domain walls are sufficiently separated. We examine this power-law scaling behavior in comparison with numerical results, shown in Fig.~\ref{fig:scaling}(b), and again find a good agreement.

\section{Braiding of Majorana states with the magnetic field}
\label{sec: braiding}

A set of $N$ localized, well separated Majorana states with second-quantized operators $\gamma_i=\gamma_i^\dagger,\ i=1,\cdots N$ that obey $\{\gamma_i,\gamma_j\}=2\delta_{ij}$ furnish, asymptotically for large $N$, a $\sqrt{2}^{N}$-dimensional fermionic Hilbert space. Fermionic operators are pairwise linear combinations $2c_{ij}:=\gamma_i+\mathrm{i}\gamma_j$, $2c^\dagger_{ij}=\gamma_i-\mathrm{i}\gamma_j$, of these Majorana operators. 
For example, two Majorana states $\gamma_1$ and $\gamma_2$, that are energetically separated from other excited states of the system, form a two-level system with eigenstates $|0\rangle$ and $|1\rangle=c_{12}^\dagger|0\rangle$, where the vacuum obeys $c_{12}|0\rangle=0$. The states $|0\rangle$ and $|1\rangle$ have Fermion parity $P_{12}:=(-1)^{c^\dagger_{12}c^{\ }_{12}}=-\mathrm{i}\gamma_1\gamma_2$ of $1$ and $-1$, respectively. An operation that changes the Fermion parity would thus change the state of this two-level system (qubit). However, parity-changing operations cannot be carried out by topological (braiding) operations. Rather, a topological qubit should be defined on a sector of constant parity. To realize this, one needs four Majorana modes to constitute a qubit. Defining the parity eigenstates of the first and second pair or Majorana modes as $|0\rangle_{12}$, $|1\rangle_{12}$ and $|0\rangle_{34}$, $|1\rangle_{34}$, respectively, we can define two states
\begin{equation}
|\bar{0}\rangle=|0\rangle_{12}\otimes |0\rangle_{34},
\qquad
|\bar{1}\rangle=|1\rangle_{12}\otimes |1\rangle_{34},
\label{eq: 4 Majonrana qbit}
\end{equation}
as a two level system. Both of these states have the same total parity $P_{12}P_{34}=+1$. States of odd parity are in principle degenerate, but cannot be accessed by the (topological) braiding operations. 

We will now show how the elementary gate operations, namely the braid
\begin{equation}
|\bar{0}\rangle\to |\bar{0}\rangle,
\qquad
|\bar{1}\rangle\to -|\bar{1}\rangle,
\end{equation}
and the $\sigma_x$ gate  
\begin{equation}
|\bar{0}\rangle\to |\bar{1}\rangle,
\qquad
|\bar{1}\rangle\to |\bar{0}\rangle.
\label{eq: sigmaz gate}
\end{equation}
can be implemented using the helical magnetic chain by an operation as simple as rotating the external magnetic field. 

\subsection{A trijunction to implement a braid between two Majorana bound states}

\begin{figure}[t]
\centering
\includegraphics[width=.47\textwidth]{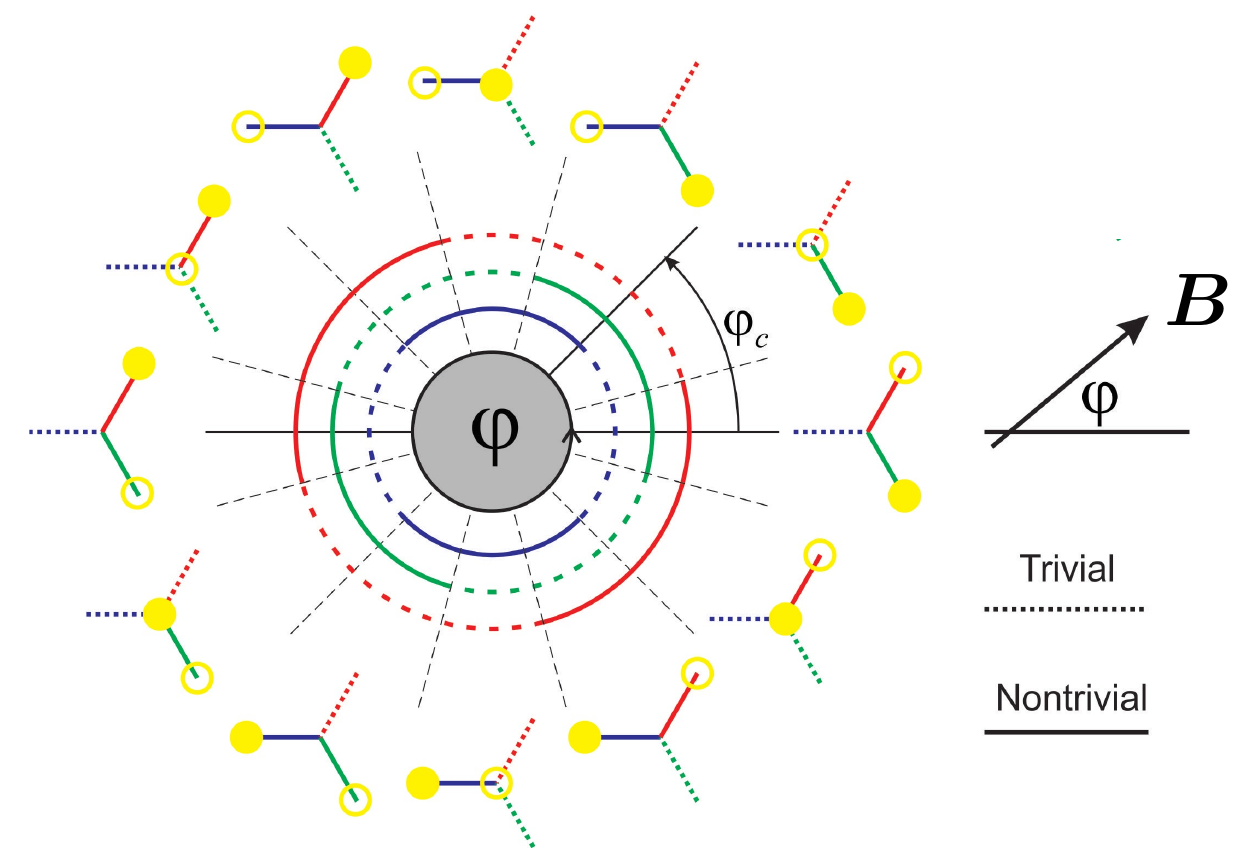}
\caption{
Implementation of the braiding operation on a qubit formed by a single pair of Majorana states in a trijunction of chains by a $2\pi$ rotation of the external magnetic field $\bs{B}$. 
}
\label{fig: Majorana trijunction}
\end{figure}

A braiding operation between two Majorana states does not change the parity of a two-Majorana qubit. To keep matters simple, we can thus demonstrate it using a single pair of Majorana fermions. The goal is to braid the two Majorana bound states in a nearly adiabatic operation. This can be achieved using a trijunction of three linear chains. If the critical angle $\varphi_0\in[\pi/6,\pi/3]$, either one or two of the three chains are in a topological state for each orientation of the magnetic field $\bs{B}$ (see Fig.~\ref{fig: Majorana trijunction}). Note that $\varphi_0$ can be conveniently tuned to the desired value by changing $|\bs{B}|$. Then, for each field orientation there are precisely 2 Majorana bound states in the trijunction. Figure.~\ref{fig: Majorana trijunction} illustrates how a $2\pi$ rotation of the magnetic field in the 1-2 plane precisely implements a braid of the two Majorana states.

\subsection{Two ellipses -- the $\sigma_x$ gate}

To implement a $\sigma_x$ gate with braiding operations, a four-Majorana qubit defined in Eq.~\eqref{eq: 4 Majonrana qbit} has to be used. It is implemented by braiding particle 1 around particle 3 (or equivalent processes). Interestingly, this can be achieved by using two overlapping ellipses with appropriately chosen critical angles $\varphi_0$ via a $2\pi$ rotation of the external magnetic field (see Fig.~\ref{fig: Majorana circle2} and the corresponding Figure in the main text). In this geometry, there are eight Majorana states present. 

For a given arrangement of ellipses, we can find out which Majorana state braids around which other one by analyzing diagrams such as Fig.~\ref{fig: Majorana circle2}(b), (d), and (f). We will explain their meaning in the following.
Consider two ellipses with equal sizes of the principal axes $p$ and $q$ that are at an angle of $\pi/2$ to each other. 
The two ellipses are defined by the two conditions
\begin{equation}
1=\frac{x_1^2}{p^2}+\frac{x_2^2}{q^2},
\qquad
1=\frac{(x_1-y_1)^2}{q^2}+\frac{(x_2-y_2)^2}{p^2},
\end{equation} 
where $\bs{y}=(y_1,y_2)$ is their relative displacement. Let us define a pair of angles $\tan\theta:=px_2/(q x_1)$ 
and $\tan\gamma:=q(x_2-y_2)/[p (x_1-y_1)]$ for the ellipses centered at $0$ and $\bs{y}$, respectively. 
The angle $\varphi$ of the tangent of the the first ellipse at $\theta$ with respect to the $x_1$-axis is determined by
\begin{equation}
\begin{split}
\tan\left(\varphi+\frac{\pi}{2}\right)
=&\frac{\mathrm{d}x_2}{\mathrm{d}x_1}\\
=&-\frac{q}{p}\frac{1}{\sqrt{1-x_1^2/p^2}}\frac{x_1}{p}\\
=&-\frac{q^2}{p^2}\frac{x_1}{x_2}\\
=&-\frac{q}{p}\frac{1}{\tan\theta},
\end{split}
\end{equation}
and analogously for the tangent to the second ellipse at $\gamma$ .
Therefore,
\begin{equation}
\tan\theta=\frac{q}{p}\tan\varphi,
\qquad
\tan\gamma=\frac{p}{q}\tan\varphi.
\label{eq: relation of the field and the normal angle}
\end{equation}
Equation~\eqref{eq: relation of the field and the normal angle} can be used to determine the following: If the critical angle for the topological phase transition of the chains is given by $\varphi_0$, then the positions of the eight Majoranas are given by the eight conditions 
\begin{equation}
\begin{split}
&\tan\theta=\frac{q}{p}\tan(\varphi\pm\varphi_0)
,\qquad
\tan\theta=\frac{q}{p}\tan(\varphi+\pi\pm\varphi_0),
\\
&\tan\gamma=\frac{q}{p}\tan(\varphi\pm\varphi_0)
,\qquad
\tan\gamma=\frac{q}{p}\tan(\varphi+\pi\pm\varphi_0).
\end{split}
\label{eq: Majorana positions on ellipses}
\end{equation}
Here, $\varphi$ is the angle of the external magnetic field with the $x_1$-axis. 
Any pair of the $4\times4$ combinations of Majorana positions $\theta$ and $\gamma$ from Eq.~\eqref{eq: Majorana positions on ellipses} can be viewed as a parametric curve in the $\theta$-$\gamma$-plane, parametrized by $\varphi\in[0,2\pi]$. These parametric curves are given for the case of two circles and a specific arrangement of two ellipses in Fig.~\ref{fig: Majorana circle2}(d) and (f), respectively.

\begin{figure}[t]
\centering
\includegraphics[width=.64\textwidth]{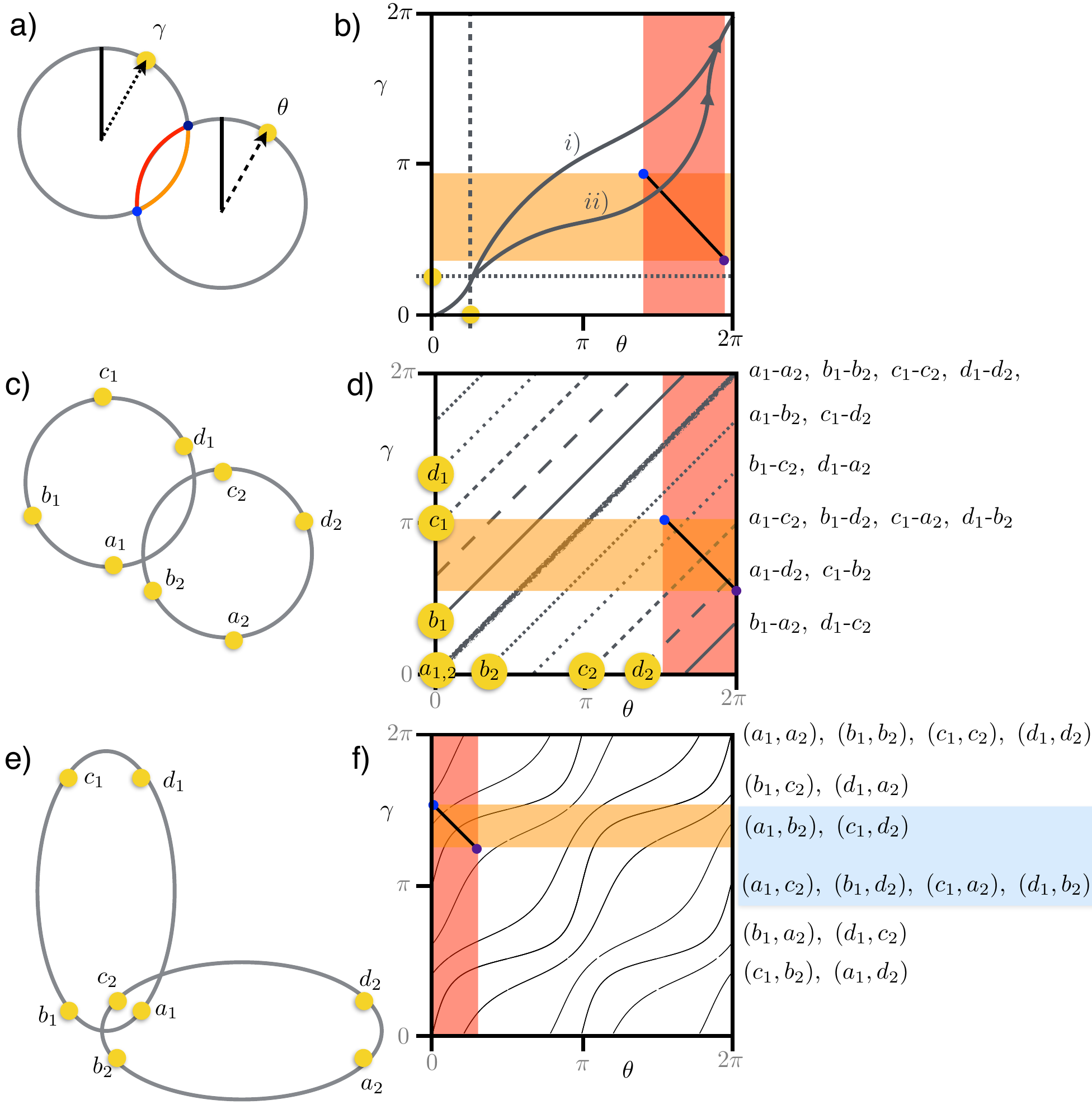}
\caption{
Implementation of a $\sigma_x$ gate with two elliptic chains via a $2\pi$ rotation of the magnetic field.
a) Two circles or ellipses are parametrized by an angle $\theta$ and $\gamma$. 
Their relative position is fixed by two intersection points (blue dots) and the information which segment of each circle is encircled by the other one (orange/red lines).
b)  $\gamma$-$\theta$-plane for the (unphysical) case of one Majorana on each circle (yellow dots). As the $\bs{B}$-field rotates, the Majoranas move along a trajectory in this plane (grey lines). If the trajectory traverses the black line connecting the two blue intersection points of the circles, the Majoranas are braided upon a $2\pi$ rotation of $\bs{B}$ [case ii)]. Otherwise, the Majoranas are not braiding [case i)].
c) The physical case of four Majoranas on each circle. 
d) Circles cannot be used to implement a $\sigma_x$ gate operation, for each Majorana fermion experiences an even number of braiding operations due to the high symmetry. This can be seen from the fact that two times an even number of trajectories intersect the black line, according to the discussion in the text. Notice that the short-dashed trajectory is 4-fold degenerate, while the dotted and the long-dahed trajectories are 2-fold degenerate, according to the Majorana pairs listed next to the trajectories. Hence, a total of $4+2\times2=2\times 4$ trajectories intersect the black line.
e) An arrangement of two ellipses that can be used to realize a $\sigma_x$ gate operation.
f) The pairs of Majoranas marked in blue are braided for this configuration of the ellipses. Again, notice that on of the trajectories that cross the black line is 4-fold degenerate and the other one is 2-fold degenerate, so that a total of $4+2=2\times3$ trajectories cross.
}
\label{fig: Majorana circle2}
\end{figure}

To see the relevance of this plot for braiding operations, let us momentarily consider the unphysical case of two circles, with one Majorana state on each circle, as depicted in Fig.~\ref{fig: Majorana circle2}(a). The relative arrangement of the two circles is determined by the interval of the angles $\theta$ and $\gamma$ that is enclosed by the other circle [red and orange segment in Fig.~\ref{fig: Majorana circle2}(a)]. In Fig.~\ref{fig: Majorana circle2}(b), this translates into a red and orange region that have an intersection spanned by the black line. Whether the two Majorana states braid upon a $2\pi$ field rotation can now be determined as follows. If the parametric curve that describes their position in the $\theta$-$\gamma$ plane as the field is rotated [the gray curve labeled i) in Fig.~\ref{fig: Majorana circle2}(b)] does not intersect the black line, the Majoranas do not braid. If the parametric curve that describes their position  in the $\theta$-$\gamma$ plane as the field is rotated intersects the black line [gray curve labeled ii) in Fig.~\ref{fig: Majorana circle2}(b)], the Majoranas braid.
  
We shall now consider the physical case of four Majorana states on each circle [i.e., an ellipse with $p=q$, see Fig.~\ref{fig: Majorana circle2}(c)]. Before returning to the more general case of two ellipses with four Majoranas each, we will convince ourselves that any arrangement of two circles is too symmetric to perform a $\sigma_x$ gate operation.
The trajectories of each pairing of a Majorana from the first circle with a Majorana from the second circle are shown as gray lines in Fig.~\ref{fig: Majorana circle2}(d). For the case of perfect circles, these are straight lines with slope 1. 
As we shall see below, to have a pair of qubits that each performs a $\sigma_x$ gate operation, it is required that two times an odd number of parametric curves intersects the black line in Fig.~\ref{fig: Majorana circle2}(d).
However, the red and orange intersection intervals cannot be chosen independently. Rather the center $\gamma'$ of the orange interval and the center $\theta'$ of the red interval are related by $\gamma'+\pi=\theta'$. One verifies that in this case the black line in Fig.~\ref{fig: Majorana circle2}(d) always embraces two times an \emph{even} number of gray curves, as these curves are arranged symmetrically around the curve defined by $\gamma+\pi=\theta$ (Notice that several parametric curves fall on top of one another in Fig.~\ref{fig: Majorana circle2}(d) and consult the figure caption for the correct counting of these degeneracies). This statement is independent of the critical angle $\varphi_0$. We conclude that two circles cannot be used to perform the desired braiding operation for a $\sigma_x$ gate.

Finally, let us consider the case of two ellipses with $p\neq q$ and take for concreteness the values $p=2q$ as well as the relative arrangement shown in Fig.~\ref{fig: Majorana circle2}(e). 
We label the Majorana states on one ellipse $a_{1}$, $b_1$, $c_1$ and $d_1$ and on the other ellipse $a_{2}$, $b_2$, $c_2$ and $d_2$.
For the ellipses, the parametric curves shown in Fig.~\ref{fig: Majorana circle2}(f) are not straight lines and the black line that characterizes the intersection of the ellipses can embrace an odd number of them.  Again, notice that several parametric curves fall on top of one another in Fig.~\ref{fig: Majorana circle2}(f). The black line is crossed by a 4-fold degenerate curve that corresponds to the Majorana pairs $a_1$-$c_2$, $a_2$-$c_1$, $b_1$-$d_2$ and $b_2$-$d_1$as well as by annother 2-fold degenerate curve is the trajectory of the pairs $a_1$-$b_2$ and $c_1$-$d_2$, making a total of $2\times 3$ trajectories crossing it. Altogether, we obtain the number of pairwise braidings upon a $2\pi$ rotation of the magnetic field as summarized in the following table.
\begin{center}
\begin{tabular}{|c|cccccccc|c|}
\hline
&$a_{1}$& $b_1$& $c_1$ & $d_1$&$a_{2}$& $b_2$& $c_2$&$d_2$&$\text{Parity of braidings}$
\\
\hline
$a_{1}$&0&1&1&1&0&1&1&0&  -1
\\
 $b_1$&1&0&1&1&0&0&0&1&  +1
 \\
  $c_1$ &1&1&0&1&1&0&0&1&  -1
  \\
   $d_1$&1&1&1&0&0&1&0&0&  +1
\\
$a_{2}$&0&0&1&0&0&1&1&1 &  +1
\\
$b_2$& 1&0&0&1&1&0&1&1&  -1
\\
$c_2$&1&0&0&0&1&1&0&1&  +1
\\
$d_2$&0&1&1&0&1&1&1&0&  -1
\\
\hline
\end{tabular}
\end{center}
We note that each Majorana in either ellipse also braids around all the other Majoranas in this ellipse. The last column of the table denotes the parity of the braiding operation on the respective Majorana, that is, the sign that the respective Majorana operator acquires.


We can now pair up the neighboring Majoranas and form a qubit from $a_1$, $b_1$ together with $a_2$, $b_2$ and a second qubit from $c_1$, $d_1$ together with $c_2$, $d_2$ in the way explained in Eq.~\eqref{eq: 4 Majonrana qbit}.
Each of these pairs of Majoranas acquires a relative phase of $-1$ during the $2\pi$ field rotation, resulting in a change of parity of the pair $P_{a_ib_i}\to-P_{a_ib_i}$, $P_{c_id_i}\to-P_{c_id_i}$, $i=1,2$. Hence, each of the two qubits changes its state from $|\bar{0}\rangle$ to $|\bar{1}\rangle$ and vice versa -- the $\sigma_x$ gate operation.


\subsection{Braiding matrices and many-body Berry phases from simulations}

\begin{figure}[t]
\centering
\includegraphics[width=.5\textwidth]{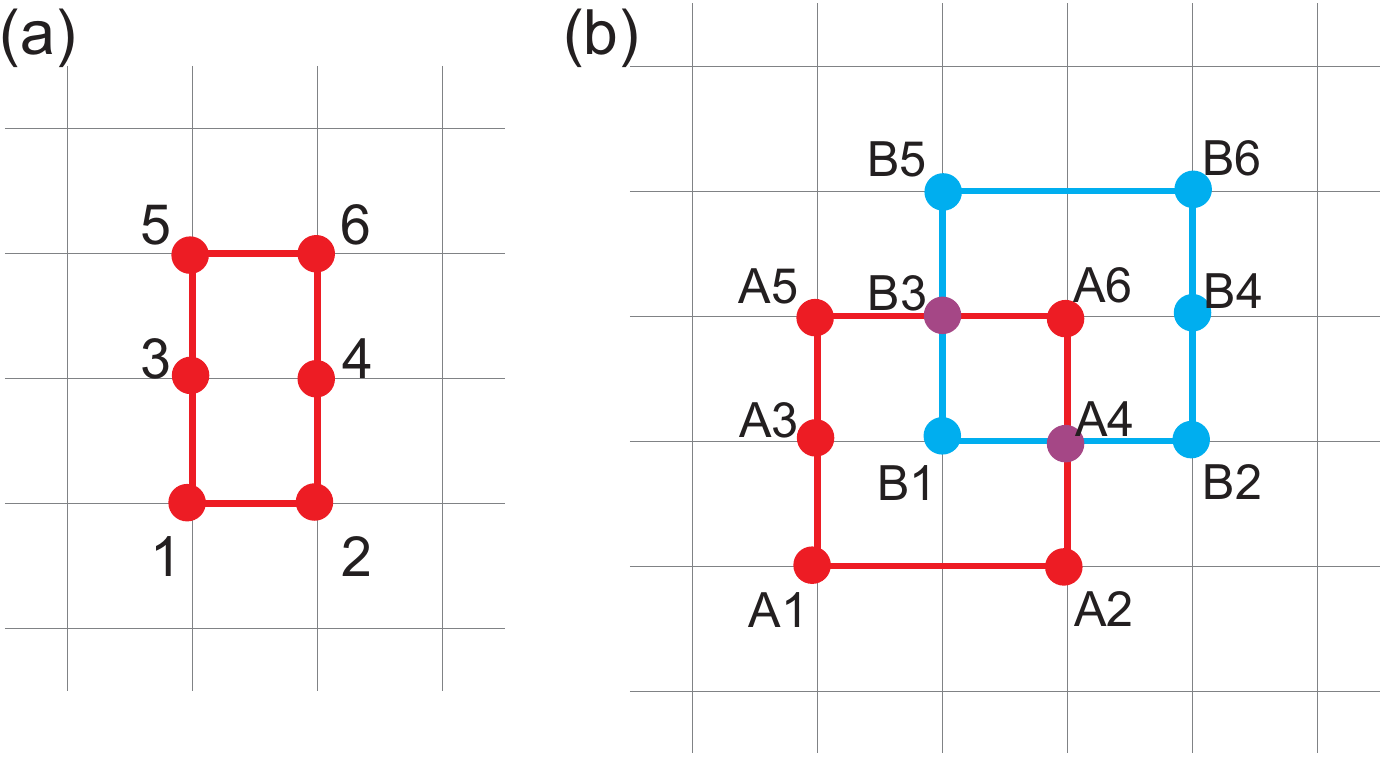}
\caption{
Minimal lattice models to simulate Majorana braiding operations in (a) a single ring and (b) two linked rings. These models are constructed from a two-dimensional lattice model for spinless $p$-wave superconductors. The construction is to pick out a small number of lattice sites and bonds (highlighted) that fulfill the intended structure. Note that in (b) we treat sites $A1$ and $A2$, as well as $B5$ and $B6$, as nearest neighbors in the original (2D) lattice.
}
\label{fig: Braiding lattice}
\end{figure}

In the previous section we derived the transformations of the states of the qubits solely based on the topology of their paths during the braid. In this section, we analyze the actual adiabatic evolution of the system under the field rotation. To that end, we reduce our original system to a simplified model in which we can extract the overall many-body Berry phase acquired by the ground states, together with the monodromy braiding matrix, upon braiding Majoranas.
The simplified models are based on a small subset of a lattice which realizes two-dimensional spinless p-wave superconductivity. To be specific, such a system is described by the following Hamiltonian
\begin{equation}
  \begin{split}
  H =&\,   \sum\limits_{\bs{n}=(n_x,n_y)} \frac{1}{2}\left[\mathbf{c}_{\bs{n}}^\dagger (i\Delta_{{\bs{n}},x}\sigma_x-t_{{\bs{n}},x}\sigma_z) \mathbf{c}_{{\bs{n}}+(1,0)} + \mathbf{c}_{\bs{n}}^\dagger (-i\Delta_{{\bs{n}},y}\sigma_y-t_{{\bs{n}},y}\sigma_z) \mathbf{c}_{{\bs{n}}+(0,1)} + \mathrm{h.c.}\right]
  \\
  &
  -
  \sum\limits_{\bs{n}=(n_x,n_y)} \mu_{\bs{n}} \mathbf{c}_{\bs{n}}^\dagger\, \sigma_z\, \mathbf{c}_{\bs{n}} 
  ,
  \end{split}
\end{equation}
where $\mathbf{c}_{\bs{n}} = (c_{\bs{n}},\;c_{\bs{n}}^\dagger)^T$ and all $\mu_{\bs{n}}$, $\Delta_{\bs{n},x/y}$ and $t_{\bs{n},x/y}$ are real. 
It is possible to base our calculation on this simplified model, as the spinless $p$-wave superconductor is a topologically equivalent effective description of our original system.

The operators $c^\dagger_{\bs{n}}$ that create spinless fermions are only defined on a subset of the sites of the lattice on which the $p$-wave superconductor resides.
The panels (a) and (b) of Fig.~\ref{fig: Braiding lattice} show two different choices of this subset, which represent a single ring and two linked rings, respectively. In each case we will abandon all the sites and bonds that are not highlighted. For simplicity, we also set $\Delta_{{\bs{n}},x}=t_{{\bs{n}},x}$ and $\Delta_{{\bs{n}},y}=t_{{\bs{n}},y}$ for each link of neighboring sites that remains. For each of these two models, we now vary slowly the set of parameters $\mathbf{p}(\lambda)$, with $\lambda$ from 0 to 1 and $\mathbf{p}(0)=\mathbf{p}(1)$, to simulate the desired operation on the Majoranas. Here, $\mathbf{p}$ represents a vector that contains all the couplings $\mu_{\bs{n}}$, $t_{{\bs{n}},x}$, and $t_{{\bs{n}},y}$ of the model. The braiding matrix associated with a specific trajectory in parameter space and a specific qubit choice can be computed numerically. By dividing the range of $\lambda$ into sufficiently small segments of length $\delta_\lambda$, the unitary transformation that represents the braiding operation in the topologically degenerate ground state manifold is well approximated by
\begin{align}
  M_{ij} = \langle GS^{(i)}_{\lambda=1} | P_{\lambda=1-\delta_\lambda}P_{\lambda=1-2\delta_\lambda}\cdots P_{\lambda=\delta_\lambda} | GS^{(j)}_{\lambda=0} \rangle,
\end{align}
if the step $\delta_\lambda$ is small enough.
Here, $GS^{(i)}_{\lambda}$ is the $i$-th ground state (with respect to a specific qubit choice) for $\mathbf{p}(\lambda)$, $GS^{(i)}_{\lambda=1}=GS^{(i)}_{\lambda=0}$ for all $i$, and $P_{\lambda}=\sum_i | GS^{(i)}_{\lambda} \rangle \langle GS^{(i)}_{\lambda} |$ is the projector to the ground state subspace at $\lambda$.

Let us first consider the single-ring case. With the labeling indicated in Fig.~\ref{fig: Braiding lattice}(a), one can verify, for example, when $\mu_3=\mu_4=t_{1x}=t_{5x}=1$ and all other parameters are zero, that there are four Majorana states, labeled by $a$, $b$, $c$ and $d$, localized at sites 2, 1, 5 and 6, respectively. Varying the parameters adiabatically according to linear interpolations between the parameter values listed in Table~\ref{tab:ring1} leads to exchanges of $a$ and $c$, and $b$ and $d$.
\begin{table}
\begin{tabular}{c|l}
\hline
nodes & non-zero parameters
\\
\hline
0 & $\mu_3,\mu_4,t_{1x},t_{5x}$ \\
1 & $t_{1x},t_{5x},t_{2y},t_{3y}$ \\
2 & $\mu_1,\mu_2,\mu_5,\mu_6,t_{1x},t_{5x}$ \\
3 & $t_{1x},t_{5x},t_{1y},t_{4y}$ \\
4 & same as 0
\\
\hline
\end{tabular}
\caption{Parameter settings at intermediate stages in an exchange operation of four Majoranas on a single ring. We only list the non-zero parameters, which are all of value 1 at the these stage. The full parameter trajectory can be obtained by linearly interpolating values between consecutive stages. \label{tab:ring1}}
\end{table}
By choosing the qubits according to two fermionic operators
\begin{align}
  f_1 = \gamma_c + i\gamma_a, \; f_2 = \gamma_b + i\gamma_d,
\end{align}
where $\gamma_{a,b,c,d}$ stand for the Majorana operators defined at time 0 for the four Majorana states, we find the braiding matrix contains only diagonal elements that correspond to the mapping
\begin{align}
  &| 00 \rangle \rightarrow e^{i\varphi}| 00 \rangle,\; | 11 \rangle \rightarrow -e^{i\varphi}| 11 \rangle, \nonumber \\
  &| 01 \rangle \rightarrow -ie^{i\varphi}| 01 \rangle,\; | 10 \rangle \rightarrow -ie^{i\varphi}| 10 \rangle,
\end{align}
under the braiding operation. Here the global phase factor $e^{i\varphi}$ contains the process-dependent many-body Berry phase. For the specific operation taken above, $e^{i\varphi} \approx e^{i 0.80 \pi}$; if we insert another intermediate stage between stages 1 and 2, for example, characterized by $\mu_1=\mu_2=\mu_5=\mu_6=t_{1x}=t_{5x}=t_{2y}=t_{3y}=1$ and otherwise 0, which only alters the specific trajectory of the parameters when linearly interpolated, but does not change the topology of the braiding operation, we find $e^{i\varphi} \approx e^{i 0.83 \pi}$.

\begin{figure}
\centering
\includegraphics[width=.8\textwidth]{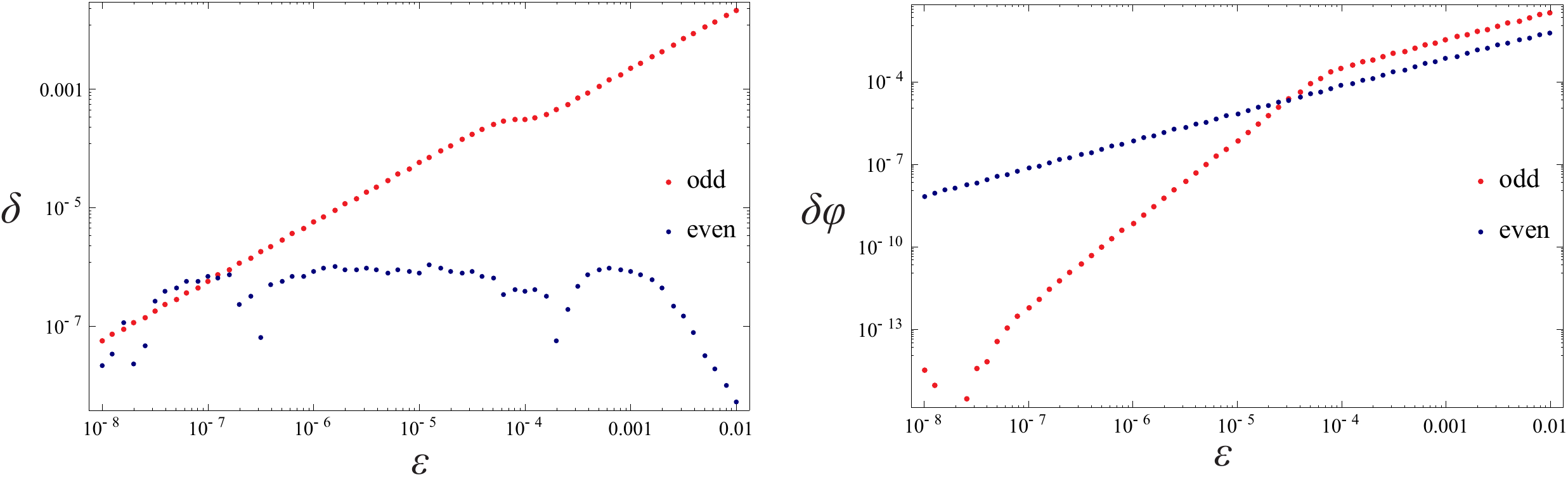}
\caption{
Errors for the operation considered in the single-ring case, as the coupling $\epsilon$ among Majorana states is increased. $\delta$ stands for magnitude of the off-diagonal elements in the braiding matrices (for the odd and the even parity sectors separately), and $\delta\varphi$ stands for the phase error in the diagonal elements.
}
\label{fig: Braiding errors}
\end{figure}

The simple and explicit model here also allows us to estimate certain kinds of errors in the operation. For instance, the parameters that we have chosen so far have an ideal property that all Majorana states are decoupled (staying at exactly zero energy) at all time during the operation. By allowing the parameters that are set to zero to be finite but small, the Majorana states become coupled and this coupling  will lead to errors in a qubit operation. Although theoretically such coupling among Majorana states can be exponentially small when their separations are large enough, for real experiments it is certainly important to estimate this kind of errors. For simplicity, we set all parameters that were previously set to 0, according to Table~\ref{tab:ring1}, to the same small value $\epsilon$, and investigate how the errors vary with $\epsilon$. Here, $\epsilon$ physically represents the energy scale of the coupling among the Majorana states. In terms of the operation we have discussed on the single ring, we consider two quantities for the errors: the magnitude $\delta$ of the off-diagonal elements in the braiding matrices, and the phase difference $\delta\varphi$ between the diagonal elements (excluding the expected relative sign change for states $| 00 \rangle$ and $| 11 \rangle$). These errors, for both parity sectors, are shown with respect to $\epsilon$ in Fig.~\ref{fig: Braiding errors}.

We now turn to the case of two linked circles. With the labeling indicated in Fig.~\ref{fig: Braiding lattice}(b), one can check that when $\mu_{A3}=\mu_{A4}=\mu_{B3}=\mu_{B4}=t_{A1x}=t_{A4x}=t_{A5x}=t_{B1x}=t_{B3x}=t_{B5x}=1$ and all other parameters are zero, there are in total eight Majorana states, labeled by $a_{1,2}$, $b_{1,2}$, $c_{1,2}$ and $d_{1,2}$ localized at sites $A2/B2$, $A1/B1$, $A5/B5$ and $A6/B6$, respectively. By varying the parameters according to linear interpolations between the parameter settings listed in Table~\ref{tab:ring2}, one realizes a Majorana braiding operation equivalent to the table that appeared in the previous section.
\begin{table}
\begin{tabular}{c|l}
\hline
nodes & non-zero parameters
\\
\hline
0 & $\mu_{A3},\mu_{A4},\mu_{B3},\mu_{B4},t_{A1x},t_{A4x},t_{A5x},t_{B1x},t_{B3x},t_{B5x}$ \\
1 & $\mu_{A4},\mu_{B3},t_{A1x},t_{A4x},t_{A5x},t_{B1x},t_{B3x},t_{B5x},t_{A3y},t_{B2y}$ \\
2 & $\mu_{A4},\mu_{A5},\mu_{A6},\mu_{B1},\mu_{B2},\mu_{B3},t_{A1x},t_{A4x},t_{A5x},t_{B1x},t_{B3x},t_{B5x}$ \\
3 & $\mu_{A4},\mu_{A5},\mu_{A6},\mu_{B1},\mu_{B3},t_{A1x},t_{A4x},t_{A5x},t_{B1x},t_{B3x},t_{B5x},t_{A4y},t_{B1y},t_{B3y}$ \\
4 & $\mu_{A4},\mu_{A6},\mu_{B1},\mu_{B3},t_{A1x},t_{A4x},t_{A5x},t_{B1x},t_{B3x},t_{B5x},t_{A2y},t_{A4y},t_{B1y},t_{B3y}$ \\
5 & $\mu_{A4},\mu_{A6},\mu_{B1},\mu_{B3},\mu_{B5},\mu_{B6},t_{A1x},t_{A4x},t_{B1x},t_{B3x},t_{B5x},t_{A2y},t_{A4y},t_{B1y}$ \\
6 & $\mu_{A1},\mu_{A2},\mu_{A4},\mu_{A6},\mu_{B1},\mu_{B3},\mu_{B5},\mu_{B6},t_{A1x},t_{B1x},t_{B3x},t_{B5x},t_{A4y},t_{B1y}$ \\
7 & $\mu_{A1},\mu_{A2},\mu_{A4},\mu_{A6},\mu_{B3},\mu_{B5},\mu_{B6},t_{A1x},t_{A5x},t_{B1x},t_{B3x},t_{B5x},t_{A4y}$ \\
8 & $\mu_{A1},\mu_{A2},\mu_{A4},\mu_{B3},\mu_{B5},\mu_{B6},t_{A1x},t_{A4x},t_{A5x},t_{B1x},t_{B3x},t_{B5x}$ \\
9 & $\mu_{A1},\mu_{A4},\mu_{B3},\mu_{B6},t_{A1x},t_{A4x},t_{A5x},t_{B1x},t_{B3x},t_{B5x},t_{A1y},t_{B4y}$ \\
$10 \rightarrow 19$ & repeat $0 \rightarrow 9$\\
20 & same as 0 \\
\hline
\end{tabular}
\caption{Parameter settings at intermediate stages in a braiding operation of Majoranas in the linked-ring case. We only list the non-zero parameters with their values set to 1 at the these stages. The full parameter trajectory can be obtained by linearly interpolating values between consecutive stages. \label{tab:ring2}}
\end{table}
By choosing the qubits according to four fermionic operators
\begin{align}
  &f_1 = \gamma_{b_1} + i\gamma_{a_1}, \; f_2 = \gamma_{c_1} + i\gamma_{d_1}, \;
  f_3 = \gamma_{c_2} + i\gamma_{d_2}, \; f_4 = \gamma_{b_2} + i\gamma_{a_2},
\end{align}
where the Majorana operators $\gamma$ are defined at time 0, we find the braiding matrix is skew-diagonal which implies
\begin{align}
  | f_4 f_3 f_2 f_1 \rangle \rightarrow e^{i\varphi} | \overline{f_4 f_3 f_2 f_1} \rangle
\end{align}
under the operation. Here, the overhead bar means a flip of all qubits (i.e., changing the fermionic occupation 0 and 1), and the (global) many-body Berry phase factor is obtained as $e^{i\varphi} \approx e^{i 0.42 \pi}$ for this specific operation. This confirms that the braiding operation indeed implements two copies of a $\sigma_x$ gate, as claimed in the previous section.

\begin{figure}
\centering
\includegraphics[width=0.85\textwidth]{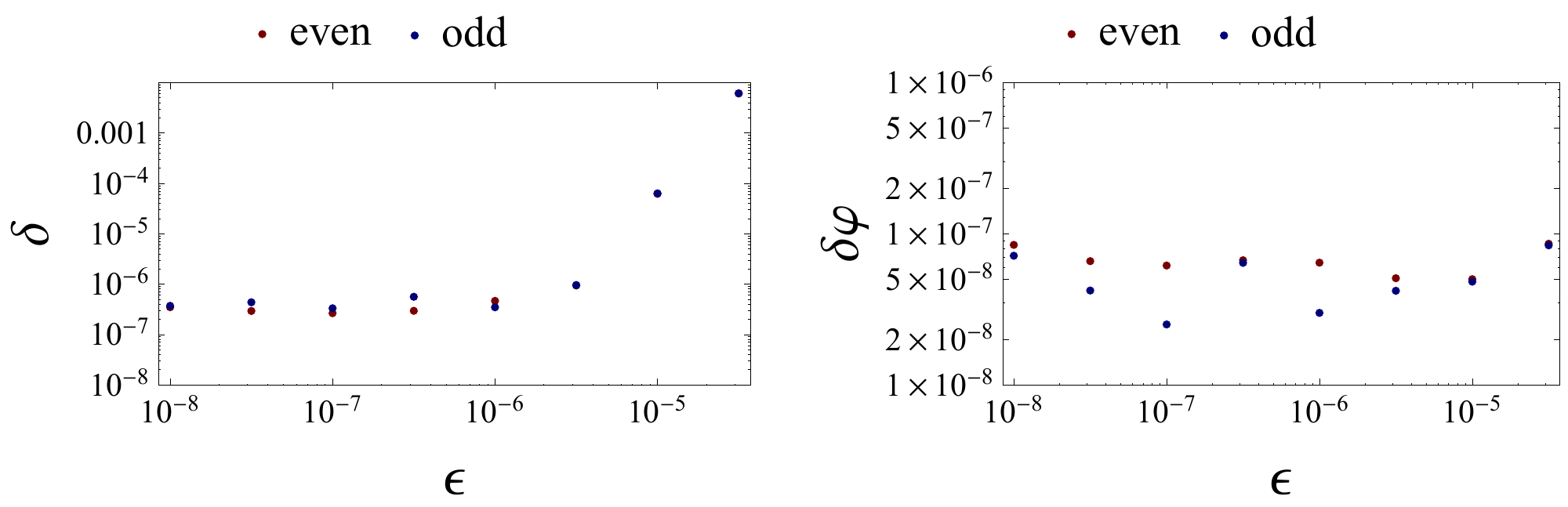}
\caption{
Errors for the operation considered in the linked-ring case, separately for the even and the odd sectors, as the coupling $\epsilon$ among Majorana states is increased. $\delta$ stands for the probability error, and $\delta\varphi$ stands for the phase error (see text for their definitions).
}
\label{fig: Two-Ring Braiding errors}
\end{figure}

Similar to the single-ring case, we can estimate the errors incurred in the current operation when the couplings among Majorana states are taken into account. Again for simplicity, we set all parameters that were previously set to 0, according to Table~\ref{tab:ring2}, to the same small value $\epsilon$, and investigate how the errors vary with $\epsilon$. Here, we define two quantities for the errors: $\delta$ is the maximum deviation from 1, of the abstract values of the skew-diagonal elements in the braiding matrix; $\delta\varphi$ is the maximum deviation of the phases of the skew-diagonal elements from the common (global) phase. We find that (see Fig.~\ref{fig: Two-Ring Braiding errors}), whereas the probability error $\delta$ increases fast beyond some threshold ($\epsilon\sim 10^{-6}$ in this case), the phase error $\delta\varphi$ is surprisingly almost independent on $\epsilon$.

As outlined in the main text, it is a technological challenge to achieve a rotating magnetic field with high enough frequency of rotation to perform this braiding operation. We propose to use current pulses of crossed conducing channels to achieve this, as outlined in Fig.~\ref{fig:CurrentPulses}. 

\begin{figure}[t]
\centering
\includegraphics[width=.7\textwidth]{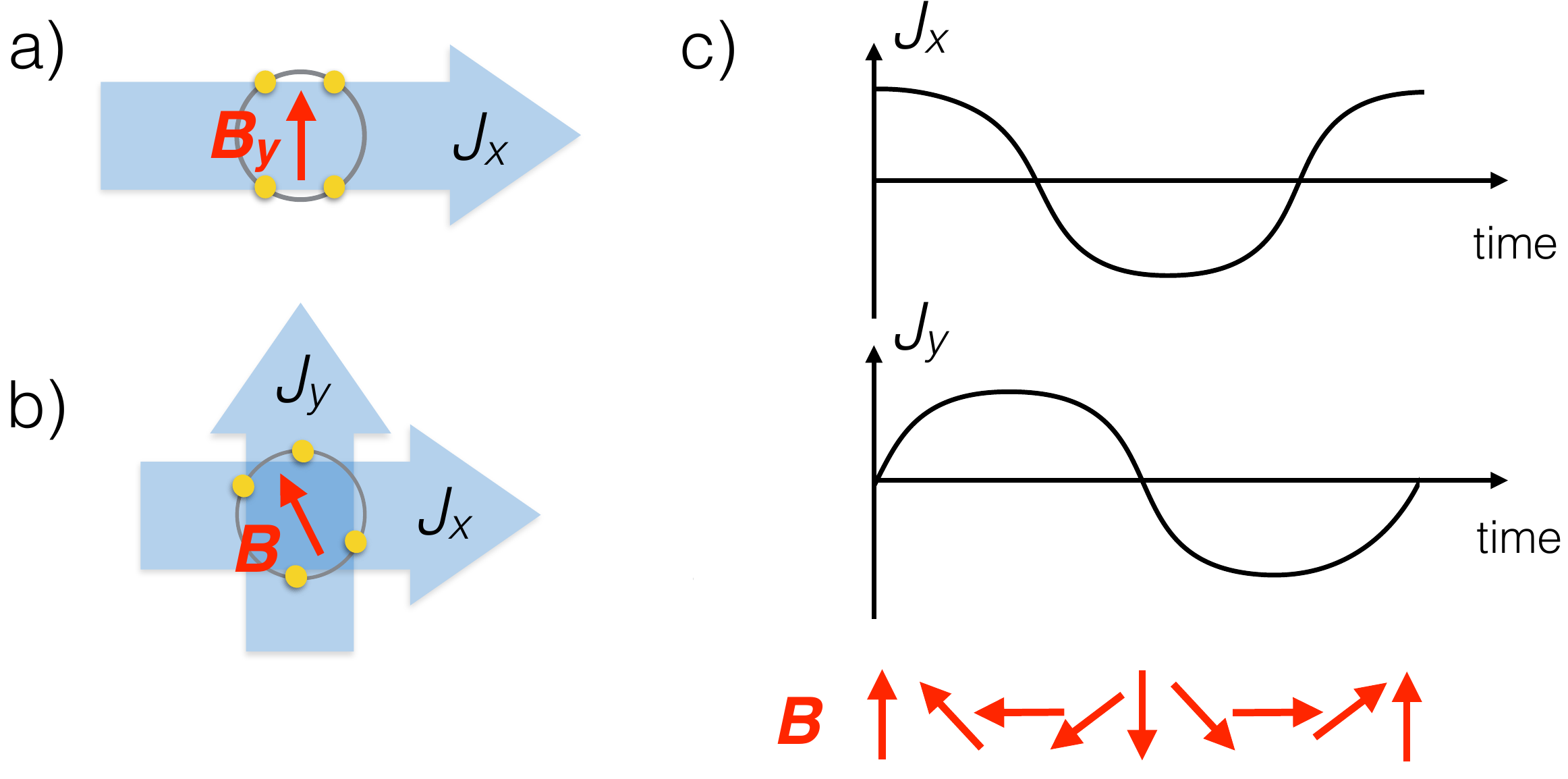}
\caption{
Possible setup for fast rotation of the magnetic field in the Majorana
necklace using current pulses.
(a) A conductor (blue) is placed below the thin superconducting film
with the chain of adatoms on top. A current $J_x$ running through the
conductor produces a magnetic field component $B_y$ perpendicular to the
current at the location of the necklace, above the conductor. (Notice
that the thin superconductor cannot screen magnetic fields which are in the
plane of the thin film.)
(b) Two perpendicular conductors with tunable currents $J_x$ and $J_y$
can be used to produce a magnetic field of arbitrary in-plane
orientation at the position of the necklace.
(c) Sinusoidal current pulses with a $\pi/2$ phase shift between $J_x$
and $J_y$ create a rotating magnetic field at the position of the necklace.
}
\label{fig:CurrentPulses}
\end{figure}

\end{widetext}

\end{document}